\documentclass{article}
\usepackage{amssymb}
\usepackage{amsmath}
\usepackage{graphicx}
\usepackage{float}
\usepackage[top=0.75in, bottom=0.75in, left=0.15in, right=0.15in]{geometry}

\begin{document}

\title{Non-Gaussian features from Excited Squeezed Vacuum State}
\author{Tang Xu-bing$^{1,2}$, Gao Fang$^{2}$, Wang Yao-xiong$^{2}$, Wu
Jian-guang$^{1}$ and Shuang Feng$^{2,3\dag }$ \\
$^{1}$School of Mathematics \& Physics Science and Engineering, \\
Anhui University of Technology, Ma'anshan 243032, China\\
$^{2}$Institute of Intelligent Machines, Chinese Academy of Sciences,\\
Hefei 230031, China\\
$^{3}$Department of Automation, University of Science \& Technology\\
\ of China, Hefei 230027, China\\
$^{\dag }$fshuang@iim.ac.cn}
\maketitle

\begin{abstract}
In this work, we introduce a non-Gaussian quantum state named excited
squeezed vacuum state (ESVS), which can be ustilized to describe quantum
light field emitted from the multiphoton quantum process occurred in some
restricted quantum systems. We investigate its nonclassical properties such
as Wigner distribution in phase space, photon number distribution, the
second-order autocorrelation and the quadrature fluctuations. By virtue of
the Hilbert-Schmidt distance method, we quantify the non-Gaussianity of the
ESVS. Due to the similar photon statistics, we examine the fidelity between
the ESVS and the photon-subtraction squeezed vacuum state (PSSVS), and then
find the optimal fidelity by monitoring the relevant parameters.
\end{abstract}

PACS numbers: 42.50.Dv, 03.67. a, 42.50.Ex

\section{Introduction}

Current research suggests that non-Gaussian states, endowed with the
qualitative role of quantum coherence or entanglement for revealling the
fascinating quantum phenomena, can preserve their nonclassicality much
better than Gaussian ones in quantum information process (QIP). Also
non-Gaussian operations become an essential ingredient for some quantum
tasks such as entanglement distillation \cite%
{Takahashi_2009_np,Dong_2008_nature} and noiseless amplication \cite%
{Xiang_2010_np}. To say the least, non-Gaussian regime has powerfully
extended to quantum information tasks, e.g. metrology \cite%
{Gilchrist_2004_jopb}, cloning \cite{Cerf_2005_prl}, communication \cite%
{Enk_2001_pra}, computation \cite{Ralph_2003_pra,Lund_2008_prl} and testing
of quantum theory \cite{Kim_2008_prl}.

In the frame of non-Gaussian mechanism, much attention has been foused on
generation schemes \cite%
{Lvovsky_2002_prl,Resch_2002_prl,Nielsen_2006_prl,Wakui_2007_oe,Bartley_2012_pra}%
, nonclassicality investigation \cite%
{XXX_2014_CPB,XXF_2014_CPB,Hu_2010_josab}, quantum protocols \cite%
{Madhok_2012_ijmpb,Datta_2008_prl,Dakic_2012_np}. In general, due to the
lack of high order nonlinearity, it is very difficult to deterministically
generate non-Gaussian states of light via optical material media. Quantum
systems with restricted dimensions have proved to be fertile ground for
discovering non-Gaussian light. For instance, in a coupled cavity-atom
system an extra resonance has been observed as well as vacuum Rabi resonance 
\cite{Schuster_2008_np}. In a photonic crystal cavity containing a strongly
coupled quantum dot, authors have discovered the photon-induced tunneling
phenomena, which is a nonclassical transmitted light \cite{Majumdar_2012_pra}%
. In circuit quantum electrodynamics (QED), the giant self-Kerr effect can
be detected by measuring the second-order correlation function and
quadrature squeezing spectrum \cite{Rebic_2009_prl}. In those coupled
microscopic quantum systems, strong interactions can generate highly
nonclassical light, which has possible uses in quantum communication \cite%
{Braunstein_2003_qf} and metrology \cite{Giovannetti_2004_science}.

One important question that arises is how to describe quantum states of
nonclassical light emitted from the restricted quantum system. Due to strong
coupling, the composite system consisting of a multi-level atom (or quantum
dot) coupled to a cavity and driven by a weak coherent field, can be
described as Jaynes-Cummings (JC) model. Quantum-optical effects can be
demonstrated in the interaction processes of photon emission and absorbtion
with atom between ground and excited states. Its energy-level structure is
discrete, so-called the dressed states \cite%
{Rempe_1987_prl,Schuster_2007_nature}, which is an overall description of
evalution. Multi-photon processes originated from quantum nonlinearity can
be monitored via the fluorescent resonance \cite%
{Maunz_2005_prl,Press_2007_prl,Hennessy_2007_nature} and the other extra
resonance \cite{Carmichael_1994_cqe}. Only considering an effective
measurement to optical field, photon-statistics methods \cite%
{Majumdar_2012_pra}, such as the second-order coherence function at time
delay zero $g^{\left( 2\right) }\left( 0\right) =\frac{\left\langle a^{\dag
}a^{\dag }aa\right\rangle }{\left\langle a^{\dag }a\right\rangle ^{2}}$ or
high order differential correlation function $C^{\left( n\right) }\left(
0\right) =\left\langle a^{\dag n}a^{n}\right\rangle -\left\langle a^{\dag
}a\right\rangle ^{n},$ are often utilized to study nonclassical
characteristics of emitting light. For the case of weak coherent incident
field, the quantum state of emitted light can be expressed as a series of
excited\ coherent states $\sum_{m}C_{m}a^{\dag m}\left\vert \alpha
\right\rangle $ or a superposition of the different Fock states $\left\vert
\psi \right\rangle =\sum_{n}C_{n}\left\vert n\right\rangle $ (due to $%
\left\vert \alpha \right\rangle =\sum_{n}\frac{\alpha ^{n}}{\sqrt{n!}}%
\left\vert n\right\rangle $) \cite{Majumdar_2012_pra}. Its photon-statistics
can exhibit similar behavior to that of an excited quantum state (e.g. $%
a^{\dag m}\left\vert \varphi \right\rangle $).

It is interesting to consider a single-mode squeezed vacuum field to be an
initial state of the incident source. Therefore, abundant nonclassicality of
emitting field can be demonstrated by studying excited squeezed vacuum state
(ESVS). In Ref.\cite{Rice_1996_pra}, considering the interaction of a
two-level atom with a squeezed vacuum, authors have calculated the
second-order intensity correlation function, the spectrum of squeezing, the
coherent spectrum and discussed nonclassical behavior of light field.
Similar issues have been addressed in Refs.\cite{Cabrillo_1995_oc,
Swain_1998_oc,Lawande_2004_jopb, An_2004_jopb,Zhang_2012_ieee,Wu_2013_pra}.
A recent experimental study found a twofold reduction of the transverse
radiative decay rate of a superconducting artificial atom coupled to
continuum squeezed vacuum \cite{Murch_2013_nature}. More attention has been
paid to investigate Wigner function and tomogram of ESVS in Ref. \cite%
{Meng_2007_pla,XXX_2014_CPB,XXF_2014_CPB,Hu_2010_josab}. In this work, we
focus on a single-mode squeezed vacuum field with any number of photon
addition, and investigate its nonclassical properties. In Sec. II, by virtue
of quantum phase space technique, we derive an analytical expression of
quasi-probability distribution Wigner function, negativity of which can
exhibit nonclassical behavior of the ESVS. And then we investigate its
photon number statistics, calculate the Mandel's $Q$ parameter, examine the
quadrature fluctuations $\left\langle \Delta X\right\rangle ,$ $\left\langle
\Delta Y\right\rangle $ and the correlation $\left\langle \Delta
X\right\rangle \left\langle \Delta Y\right\rangle ,$ which can be measured
outside the cavity by using a homodyne detection with a controllable phase.
In Sec. III, we evaluate its non-Gaussianity via the Hilbert-Schmidt
distance method \cite{Genoni_2007_pra}. As results, we shall study how the
photon number modulation affects the non-Gaussianity of the ESVS and provide
a guide to enhance non-Gaussianity of a desired quantum state. It is found
that photon number distribution of the ESVS is similar to that of the
photon-subtraction squeezed vacuum state (PSSVS). Fidelity between the ESVE
and the PSSVS is obtained and the optimal fidelity has been discussed in
Sec. IV. We end with the main conclusions of our work.

\section{Nonclassical Properties Investigation of the ESVS}

As is well known, a single mode squeezed field is an approximation with a
superposition of all even number photon states, i.e., 
\begin{equation}
S\left( r\right) \left\vert 0\right\rangle =\exp \left[ \frac{r}{2}\left(
a^{\dag 2}-a^{2}\right) \right] \left\vert 0\right\rangle =\frac{1}{\sqrt{%
\cosh r}}\sum_{n=0}\left( -1\right) ^{n}\frac{\sqrt{\left( 2n\right) !}}{%
2^{n}n!}\tanh ^{n}r\left\vert 2n\right\rangle ,  \label{sq}
\end{equation}%
in which $r$ is the squeezing parameter. Adding one single photon to a weak
squeezed field can be described by 
\begin{equation}
a^{\dag }S\left( r\right) \left\vert 0\right\rangle \rightarrow
\sum_{n=0}\left( -1\right) ^{n}\frac{\sqrt{\left( 2n+1\right) !}}{2^{n}n!}%
\tanh ^{n}r\left\vert 2n+1\right\rangle ,  \label{csq}
\end{equation}%
which is a superposition of all odd number photon states. Theoretically, by
repeating ($n$ times) application of the photon creation operator $a^{\dag }$
on the squeezed vacuum field, we can obtain the excited squeezed vacuum
state (ESVS) $a^{\dag n}S\left( r\right) \left\vert 0\right\rangle .$ For
the case of small value of $n$, the ESVS can be generated via the
spontaneous parametric down-conversion (SPDC) occured in nonlinear optical
media or the multiphoton processes in above-mentioned restricted quantum
systems. Thus its density operator reads 
\begin{equation}
\rho \left( r,n\right) =C_{n}^{-1}a^{\dag }{}^{n}S\left( r\right) \left\vert
0\right\rangle \left\langle 0\right\vert S^{-1}\left( r\right) a^{n}\equiv
C_{n}^{-1}a^{\dag }{}^{n}\rho \left( r\right) a^{n},  \label{do}
\end{equation}%
in which $\rho \left( r\right) \equiv S\left( r\right) \left\vert
0\right\rangle \left\langle 0\right\vert S^{-1}\left( r\right) $ denotes the
squeezed vacuum field, $C_{n}=Tr\left[ \rho \left( r,n\right) \right]
=n!\cosh ^{n}rP_{n}\left( \cosh r\right) $ is a normalized constant. $%
P_{n}\left( \cosh r\right) $\ is the expression of the Legendre polynomials
and this result has also obtained in Ref.\cite{Meng_2007_pla}.

\subsection{Quasi-probability Distribution: Wigner Function}

In order to interview the nonclassical properties of a quantum light fields,
the Wigner function, although not positive definition in general, provides a
closely parallel interpretation as a probability distribution function.
Based on the Weyl's mapping rule \cite{Weyl_1927_phys,Fan_2003_jopb_R}, the
classical correspondence of density operator $\rho $ is just the Wigner
function, namely%
\begin{equation}
\rho =\int \int_{-\infty }^{\infty }dqdp\Delta \left( q,p\right) W\left(
q,p\right)  \label{weyl}
\end{equation}%
or%
\begin{equation}
W\left( q,p\right) =Tr\left[ \rho \Delta \left( q,p\right) \right] ,
\label{wf}
\end{equation}%
$W\left( q,p\right) $ is the Wigner function of $\rho $ and $\Delta \left(
q,p\right) $ denotes the Wigner operator, defined in the coordinate
representation $\left\vert q\right\rangle $ as 
\begin{equation}
\Delta \left( p,q\right) =\int_{-\infty }^{\infty }\frac{dv}{2\pi }%
e^{ipv}\left\vert q+\frac{v}{2}\right\rangle \left\langle q-\frac{v}{2}%
\right\vert .  \label{wo}
\end{equation}%
Noting that $\alpha =\frac{1}{\sqrt{2}}\left( q+ip\right) ,$ we can see $%
W\left( \alpha ,\alpha ^{\ast }\right) =Tr\left[ \rho \Delta \left( \alpha
,\alpha ^{\ast }\right) \right] ,$ where 
\begin{eqnarray}
\Delta \left( \alpha ,\alpha ^{\ast }\right) &=&\int \frac{d^{2}z}{\pi ^{2}}%
\left\vert \alpha +z\right\rangle \left\langle \alpha -z\right\vert
e^{\alpha z^{\ast }-\alpha ^{\ast }z}=\frac{1}{\pi }\colon \exp \left[
-2\left( a^{\dagger }-\alpha ^{\ast }\right) \left( a-\alpha \right) \right]
\colon  \notag \\
&=&\frac{1}{2}%
\begin{array}{c}
\dot{\vdots}%
\end{array}%
\delta \left( \alpha -a\right) \delta \left( \alpha ^{\ast }-a^{\dagger
}\right) 
\begin{array}{c}
\dot{\vdots}%
\end{array}%
,  \label{7}
\end{eqnarray}%
$\left\langle z\right\vert =\left\langle 0\right\vert \exp \left[ -\frac{1}{2%
}|z|^{2}+z^{\ast }a\right] $ is the Glauber coherent state \cite%
{Glauber_1963_pr}, the symbols ' $\colon \colon $' and '$%
\begin{array}{c}
\dot{\vdots}%
\end{array}%
\begin{array}{c}
\dot{\vdots}%
\end{array}%
$'denote the normal ordering and Weyl ordering, repectively. In particular,
a unitary operator (e.g. $S$ with its identities $SaS^{-1}=\mu a+\nu
a^{\dagger },$ $Sa^{\dagger }S^{-1}=\sigma a+\tau a^{\dagger })$ can 'run
across' the 'border' of '$%
\begin{array}{c}
\dot{\vdots}%
\end{array}%
\begin{array}{c}
\dot{\vdots}%
\end{array}%
$' and directly transforms bosonic operators, i.e. 
\begin{equation}
SF\left( a^{\dagger },a\right) S^{-1}=F\left( \mu a+\nu a^{\dagger },\sigma
a+\tau a^{\dagger }\right) =%
\begin{array}{c}
\dot{\vdots}%
\end{array}%
f\left( \mu a+\nu a^{\dagger },\sigma a+\tau a^{\dagger }\right) 
\begin{array}{c}
\dot{\vdots}%
\end{array}%
,  \label{invariance}
\end{equation}%
which also named the Weyl ordering invariance under similarity
transformations \cite{Fan_2006_ap}. Substituting Eq.(\ref{do}) into Eq.(\ref%
{wf}), we can derive the Wigner function of the ESVS, namely%
\begin{eqnarray}
W\left( \alpha ,\alpha ^{\ast }\right) &=&\frac{1}{P_{n}\left( \cosh
r\right) }\left( \frac{\sinh r}{2}\right) ^{n}\exp \left[ -2\left\vert
\alpha \cosh r-\alpha ^{\ast }\sinh r\right\vert ^{2}\right]  \notag \\
&&\times \sum_{m=0}^{n}\binom{n}{m}\left( -2\coth r\right) ^{m}\left\vert
H_{n-m}\left[ -i\sqrt{\frac{2}{\tanh r}}\left( \alpha \cosh r-\alpha ^{\ast
}\sinh r\right) \right] \right\vert ^{2},  \label{wigner}
\end{eqnarray}%
where $H_{n}\left( x\right) $ is the Hermite polynomials defined as 
\begin{equation}
H_{n}\left( x\right) =x^{n}\sum_{k=0}^{\left[ n/2\right] }\frac{n!}{%
2^{2k}\left( k!\right) ^{2}\left( n-2k\right) !}\left( 1-\frac{1}{x^{2}}%
\right) ^{k}.  \label{hermite}
\end{equation}%
As we can see in Eq.(\ref{wigner}), the exponential function $\exp \left[
-2\left\vert \alpha \cosh r-\alpha ^{\ast }\sinh r\right\vert ^{2}\right] $
is a discription of Gaussian distribution and the Wigner function $W\left(
\alpha ,\alpha ^{\ast }\right) $ of the ESVS is a sum of products of the
Hermite-Gaussian functions. Details in the derivation of Eq.(\ref{wigner})
has been shown in Appendix A.

\begin{figure}[h]
\centering
\includegraphics[scale=0.50]{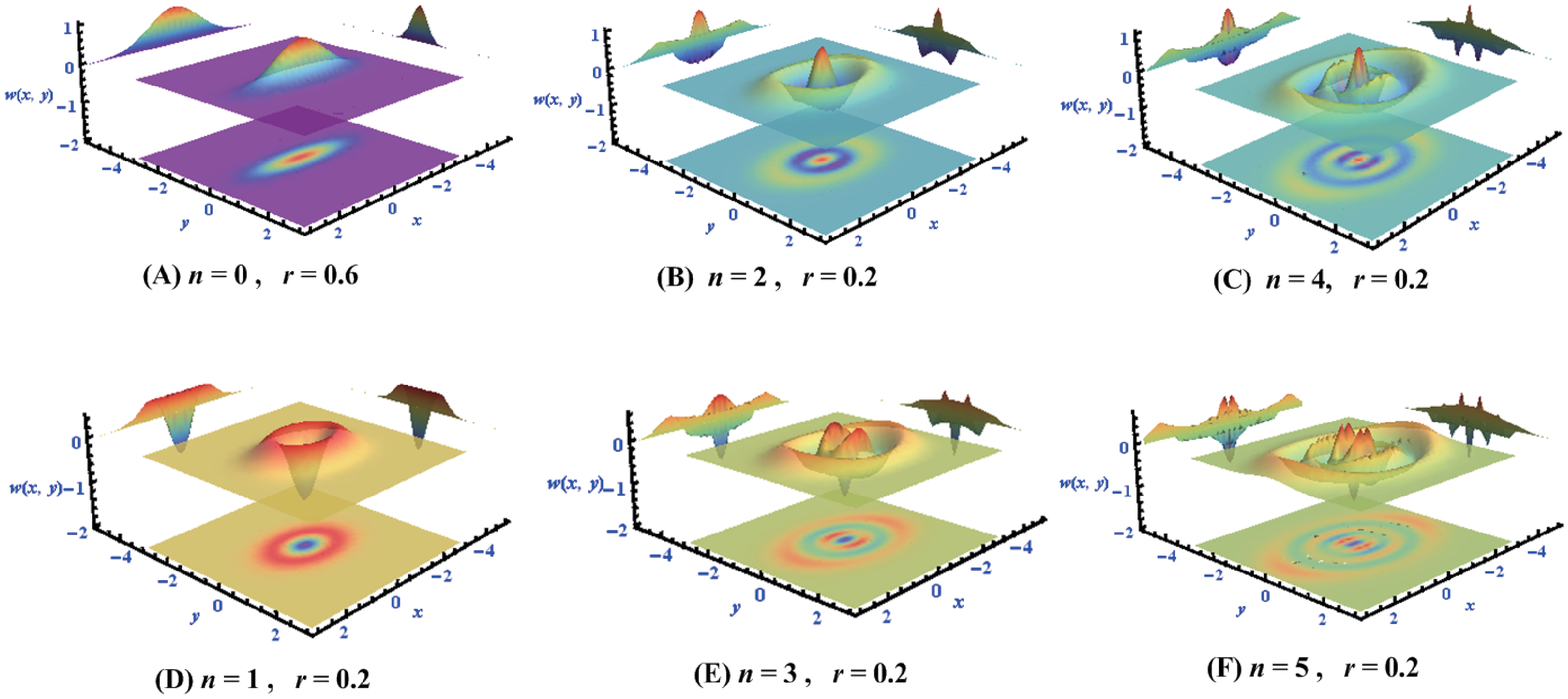}
\caption{\textsf{Wigner functions of the ESVS with fixed $n$ and $r$ : (A). $%
n=0,$ $r=0.6,$ (B). $n=2,$ $r=0.2,$ (C). $n=4,$ $r=0.2,$ (D). $n=1,$ $r=0.2,$
(E). $n=3,$ $r=0.2,$ (F). $n=5,$ $r=0.2.$ In the top row, Figure 1(A)
exhibits a squeezed vacuum state, and in Figure 1(B)--1(C) adding even
photon shows a a smaller positive value at the center. In the bottom row
(Figure 1(D)--1(F)), adding odd photon leads to a negative value at the
center of phase space. More added-photon means a much more complex
distribution.}}
\end{figure}

Figure 1 shows the quantum non-Gaussian states can be preapred by adding
photon to a weak light field. At low intensities of the squeezed vacuum
field (see Figure 1(A)), the variance of photon-addition can exhibit
different nonclassical features (negative distribution). (B) and (C)
describe the even photon-addition states, odd photon-addition states in (D),
(E) and (F). Indeed, the squeezing parameter $r$ dominates the quadrature
distribution in the directions of $-X$ and $-Y$. Here the values of $r$ are $%
0.6$ in (A) and $0.2$ in (B)-(F), respectively.

\subsection{Photon Number Distribution}

In Eq.(\ref{do}), noting that $\left\vert 0\right\rangle \left\langle
0\right\vert =\colon \exp \left( -a^{\dag }a\right) \colon ,$ we have 
\begin{equation}
\rho \left( r\right) \equiv S\left( r\right) \left\vert 0\right\rangle
\left\langle 0\right\vert S^{-1}\left( r\right) =\frac{1}{\cosh r}\colon
\exp \left[ \frac{1}{2}a^{\dag 2}\tanh r+\frac{1}{2}a^{2}\tanh r-a^{\dag }a%
\right] \colon  \label{3}
\end{equation}%
Photon number distribution (PND) of the ESVS can be defined via the
probability of finding $m$ photons, namely 
\begin{eqnarray}
P_{m}\left( r,n\right) &=&\left\langle m\right\vert \rho \left( r,n\right)
\left\vert m\right\rangle  \notag \\
&=&C_{n}^{-1}\frac{1}{\cosh r}\frac{m!}{\left( m-n\right) !}\left\langle
m-n\right\vert \colon \exp \left[ \frac{1}{2}a^{\dag 2}\tanh r+\frac{1}{2}%
a^{2}\tanh r-a^{\dag }a\right] \colon \left\vert m-n\right\rangle .
\label{4}
\end{eqnarray}%
And using $\left\vert n\right\rangle =\frac{1}{\sqrt{n!}}\left. \frac{d^{n}}{%
d\alpha ^{n}}\left\vert \alpha \right\rangle \right\vert _{\alpha =0}$ , $%
\left\langle \beta \right. \left\vert \alpha \right\rangle =\exp \left(
\alpha \beta ^{\ast }\right) $ and the differential form of the Hermite
polynomials 
\begin{equation}
H_{n}\left( x\right) =\left( -1\right) ^{n}e^{x^{2}}\frac{d^{n}}{dx^{n}}%
e^{-x^{2}},  \label{5}
\end{equation}%
we have%
\begin{eqnarray}
P_{m}\left( r,n\right) &=&C_{n}^{-1}\frac{1}{\cosh r}\frac{m!}{\left[ \left(
m-n\right) !\right] ^{2}}\left. \frac{d^{m-n}}{d\beta ^{\ast m-n}}\frac{%
d^{m-n}}{d\alpha ^{m-n}}\exp \left[ \frac{1}{2}\beta ^{\ast 2}e^{-i\theta
}\tanh r+\frac{1}{2}\alpha ^{2}e^{i\theta }\tanh r\right] \right\vert
_{\alpha ,\beta ^{\ast }=0}  \notag \\
&=&C_{n}^{-1}\frac{1}{\cosh r}\frac{m!}{\left[ \left( m-n\right) !\right]
^{2}}\left( -\frac{\tanh r}{2}\right) ^{m-n}\left[ H_{m-n}\left( 0\right) %
\right] ^{2},  \label{photon}
\end{eqnarray}%
where $C_{n}\ $is given in Eq.(\ref{do}). For the case of $x=0,$ we can see 
\begin{equation}
H_{2n}\left( 0\right) =\left( -1\right) ^{n}\frac{\left( 2n\right) !}{n!},%
\text{ \ \ \ }H_{2n+1}\left( 0\right) =0.  \label{even}
\end{equation}%
Therefore, the results of $P_{m}\left( r,n\right) $ can be decided by the
value of $m-n.$ Noting a differential identity for the Legendre polynomials,
i.e. 
\begin{equation}
\left. \frac{\partial {}^{2m}}{\partial t{}^{m}\partial \tau {}^{m}}\exp %
\left[ -t{}^{2}-\tau {}^{2}+\frac{2xt\tau }{\sqrt{x^{2}-1}}\right]
\right\vert _{t,\tau =0}=\frac{2^{m}m!}{\left( x^{2}-1\right) ^{m/2}}%
P_{m}\left( x\right) ,  \label{6}
\end{equation}%
we can rewrite Eq.(\ref{photon}) as 
\begin{eqnarray}
P_{m}\left( r,n\right) &=&\frac{C_{n}^{-1}}{\cosh r}\frac{m!}{\left[ \left(
m-n\right) !\right] ^{2}}\left( -\frac{\tanh r}{2}\right) ^{m-n}\left. \frac{%
d^{m-n}}{d\beta ^{\ast m-n}}\frac{d^{m-n}}{d\alpha ^{m-n}}\exp \left[ -\beta
^{\ast 2}-\alpha ^{2}\right] \right\vert _{\alpha ,\beta ^{\ast }=0}  \notag
\\
&=&\frac{C_{n}^{-1}}{\cosh r}\left( -\frac{\tanh r}{2}\right) ^{m-n}\frac{m!%
}{\left( m-n\right) !}\frac{2^{m-n}}{\left( -1\right) ^{(m-n)/2}}%
P_{m-n}\left( 0\right) .  \label{12}
\end{eqnarray}

\begin{figure}[h]
\centering
\includegraphics[scale=0.50]{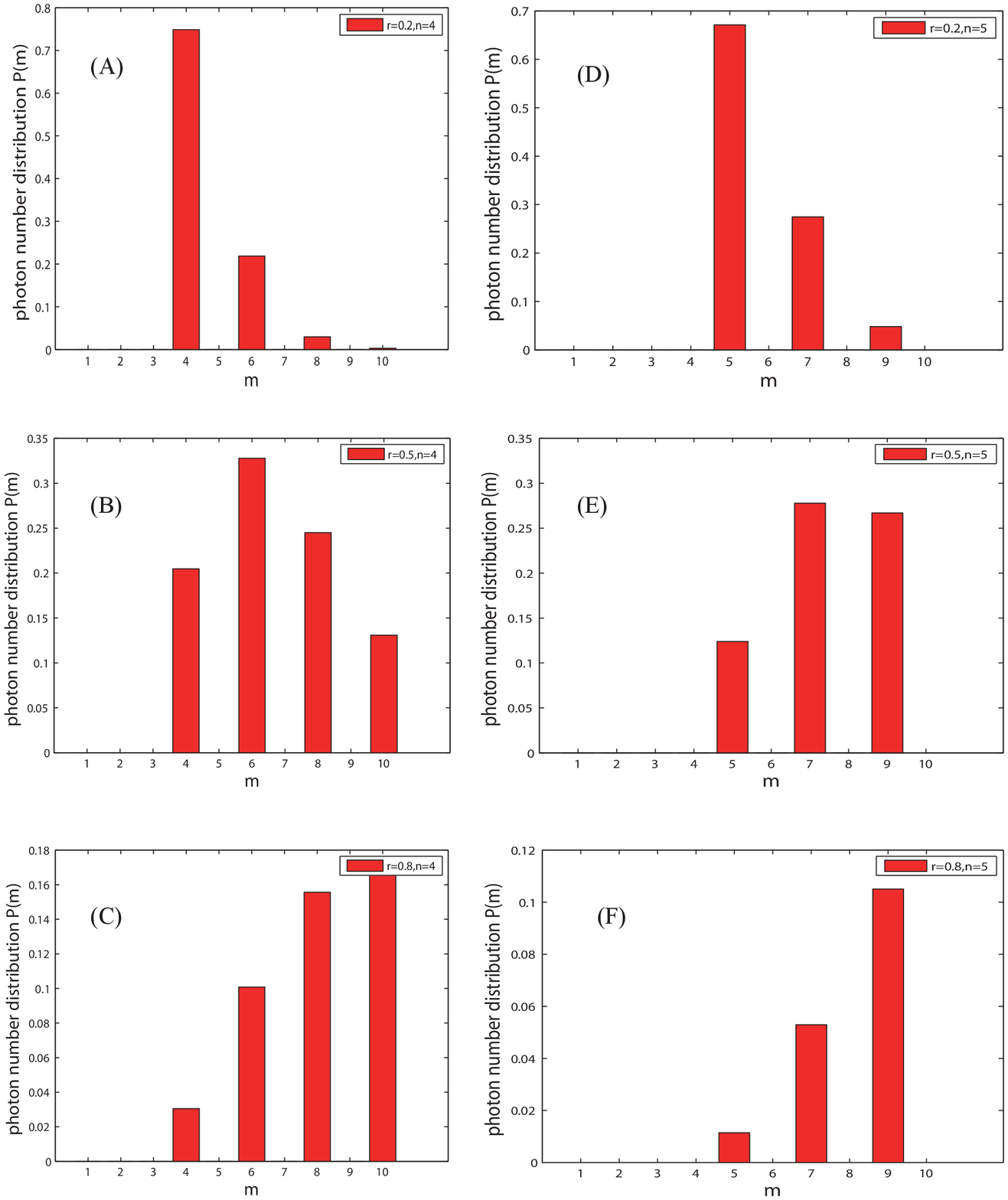}
\caption{\textsf{Photon number distributions of the ESVE with fixed
added-photon number $n.$ Enhancing squeezing leads to the transition of
photon number probability distribution from the left column (Figure
2(A)--2(C)) with $n=4$ and $r=0.2,$ $0.5,$ $0.8,$ to the right (Figure
2(D)--2(F)) with $n=5$ and $r=0.2,$ $0.5,$ $0.8$.}}
\end{figure}

Indeed, the squeezed vacuum field is close to the even number photon
superposition states. For the case of an odd photon-addition, the final
state can be described by an odd photon superposition states, for details
see Eq.(\ref{spssvs}) in Sec. IV. This point has also been shown on the
right column of Figure 2. When rising value of the squeezing parameter $r,$
we can see that the probability distribution has gradually transited towards
to the larger photon number distribution (see from figure 2(D) to figure
2(E)). If adding even photon, this distribution can be described by an even
photon superposition states, as you can see on the left column of Figure 2.
For the case of an odd photon-addition, the similar results can also be
obtained by enhancing the squeezing $r$ (see from Figure 2(A) to 2(C)).

\subsection{Mandel's $Q$ Parameter}

In order to show the occupation of photon number distribution, the above
mentioned second-order coherence function at time delay zero $g^{\left(
2\right) }\left( 0\right) $ is often utilized to measure the nonclassicality
of quantum light. Alternatively, Ref.\cite{Mandel_1979_ol} defined the
Mandel's $Q$ parameter, 
\begin{equation}
Q=\frac{\left\langle a^{\dag 2}a^{2}\right\rangle }{\left\langle a^{\dag
}a\right\rangle }-\left\langle a^{\dag }a\right\rangle =\frac{\left\langle 
\hat{n}^{2}\right\rangle -\left\langle \hat{n}\right\rangle ^{2}}{%
\left\langle \hat{n}\right\rangle }-1=\left\langle \hat{n}\right\rangle %
\left[ g^{\left( 2\right) }\left( 0\right) -1\right]  \label{11}
\end{equation}%
to characterize nonclassicality with negative values indicating a
sub-Poissonian statistics in resonance fluorescence. The minimal value $Q=-1$
indicates the photon number states and $-1\leq Q<0$ can be interpreted as a
nonclassical probability distribution. $Q=0$ means a Poissonian photon
number statistics, which is mostly close to the classical probability
distribution (e.g. coherent state). For the case of $Q>0,$ light field is
considered to be the super-Poissonian distribution. From Eq. (\ref{do}) and
noting that 
\begin{eqnarray}
\left\langle a^{\dag }a\right\rangle &=&\left\langle \lambda ,n\right\vert
a^{\dag }a\left\vert \lambda ,n\right\rangle =C_{n}^{-1}\left\langle
0\right\vert S^{-1}\left( \lambda \right) a{}^{n}a^{\dag }aa^{\dag
}{}^{n}S\left( \lambda \right) \left\vert 0\right\rangle =\frac{C_{n+1}-C_{n}%
}{C_{n}},  \notag \\
\left\langle a^{\dag 2}a^{2}\right\rangle &=&\left\langle \lambda
,n\right\vert a^{\dag 2}a^{2}\left\vert \lambda ,n\right\rangle
=C_{n}^{-1}\left\langle 0\right\vert S^{-1}\left( \lambda \right)
a{}^{n}a^{\dag 2}a^{2}a^{\dag }{}^{n}S\left( \lambda \right) \left\vert
0\right\rangle =\frac{C_{n+2}}{C_{n}}-4\frac{C_{n+1}}{C_{n}}+2,  \label{10}
\end{eqnarray}%
the Mandel's Q parameter is given by%
\begin{equation}
Q=\frac{\left\langle a^{\dag 2}a^{2}\right\rangle }{\left\langle a^{\dag
}a\right\rangle }-\left\langle a^{\dag }a\right\rangle =\frac{C_{n+2}-2C_{n}%
}{C_{n+1}-C_{n}}-\frac{C_{n+1}}{C_{n}}-3,  \label{mandel}
\end{equation}%
where coeffcient $C_{n+j}=\left( n+j\right) !\cosh ^{n+j}rP_{n+j}\left(
\cosh r\right) $ has defined in Eq.(\ref{do}). As a function of parameters $%
n $ and $r$, the Mandel's $Q$ \ has been shown in Figure 3(A). From the top
to bottom, the values of $n$ are fixed with $0$, $1$ and $6$. For the case
of $n=0,$ $Q\geqslant 1$ means the squeezing vacuum field has
super-Poissonian photon statistics (see orange line in Figure 3(A)). Besides 
$n=0,$ the ESVS is sub-Poissonian, Poissonian and super-Poissonian with the
different ranges of squeezing parameter $r\ $(see red line and dashed line
in\ Figure 3(A)). If $n\neq 0,$ the ESVS is always a nonclassical state, we
can see that $Q>0$ does not mean that the state is classical. This point can
be confirmed by taking $n=1$ and $6$ in Figure 3(A). Obviously, at lower
intensity squeezing, the ESVS is sub-Poissonian photon statistics, while
high strength means the super-Poisson.

\subsection{Quadrature Fluctuations}

The ESVS is quadrature squeezed and can yield an $\hat{X}$ quadrature
variance $\Delta \hat{X}$ above the standard quantum limit (SQL) at the cost
of the other $\hat{Y}$ quadrature variance $\Delta \hat{Y}$ below the SQL.
In order to investigate its squeezing behaviour, firstly we calculate the
expected value of a general operator $a^{k}a^{\dag }{}^{l}$ under the ESVS,

\begin{eqnarray}
\left\langle r,n\right\vert a^{k}a^{\dag }{}^{l}\left\vert \lambda
,n\right\rangle &=&C_{n}^{-1}\left\langle 0\right\vert S^{-1}\left( r\right)
a^{n}a^{k}a^{\dag }{}^{l}a^{\dag }{}^{n}S\left( r\right) \left\vert
0\right\rangle  \notag \\
&=&C_{n}^{-1}\left\langle 0\right\vert \left( a\cosh r+a^{\dag }\sinh
r\right) ^{n+k}\left( a^{\dag }\cosh r+a\sinh r\right) ^{n+l}\left\vert
0\right\rangle  \label{9}
\end{eqnarray}%
and we can obtain%
\begin{eqnarray}
\left\langle \lambda ,n\right\vert a^{k}a^{\dag }{}^{l}\left\vert \lambda
,n\right\rangle &=&C_{n}^{-1}\left( -i\sqrt{\frac{\sinh 2r}{4}}\right)
^{k}\left( i\sqrt{\frac{\sinh 2r}{4}}\right) ^{l}\left( \frac{\sinh 2r}{4}%
\right) ^{n}  \notag \\
&&\times \sum_{m=0}^{n+k}\dbinom{n+k}{m}\frac{\left( -1\right)
^{n+k-m}2^{m}\left( n+l\right) !}{\left( n+l-m\right) !}\left( \coth
r\right) ^{m}H_{n+l-m}\left( 0\right) H_{n+k-m}\left( 0\right) .
\label{general}
\end{eqnarray}%
From Eq.(\ref{even}), the conditions for existance of non-zero value as you
can see in Eq.(\ref{general}) that $n+l-m$ and $n+k-m$\ must also take even.
The above derivation for details have shown in Appendix B. For the cases of $%
k=1,$ $l=0$ and $k=0,$ $l=1,$\ we can see one of $n-m$ and $n+1-m$\ must be
odd. Thus we have%
\begin{eqnarray}
\left\langle r,n\right\vert a\left\vert r,n\right\rangle
&=&-iC_{n}^{-1}\left( \frac{\sinh 2r}{4}\right) ^{n+\frac{1}{2}%
}\sum_{m=0}^{n+1}\dbinom{n+1}{m}\frac{\left( -1\right) ^{n+1-m}2^{m}n!}{%
\left( n-m\right) !}\left( \coth r\right) ^{m}H_{n-m}\left( 0\right)
H_{n+1-m}\left( 0\right) =0  \notag \\
\left\langle r,n\right\vert a^{\dag }\left\vert r,n\right\rangle
&=&iC_{n}^{-1}\left( \frac{\sinh 2r}{4}\right) ^{n+\frac{1}{2}%
}\sum_{m=0}^{n+1}\dbinom{n+1}{m}\frac{\left( -1\right) ^{n+1-m}2^{m}n!}{%
\left( n-m\right) !}\left( \coth r\right) ^{m}H_{n-m}\left( 0\right)
H_{n+1-m}\left( 0\right) =0  \label{one}
\end{eqnarray}%
For the cases of $k=2,$ $l=0$ and $k=0,$ $l=2,$ we have\ 
\begin{equation}
\left\langle r,n\right\vert a^{2}\left\vert r,n\right\rangle =\left\langle
r,n\right\vert a^{\dag 2}\left\vert r,n\right\rangle =M\left( n,r\right)
\label{two}
\end{equation}%
where $M\left( n,r\right) \equiv -C_{n}^{-1}\left( \frac{\sinh 2r}{4}\right)
^{n+1}\sum_{m=0}^{n+2}\dbinom{n+2}{m}\frac{2^{m}n!}{\left( n-m\right) !}%
\left( \coth r\right) ^{m}H_{n-m}\left( 0\right) H_{n+2-m}\left( 0\right) .$
When taking $k=l=1,$ we have $\left\langle r,n\right\vert aa^{\dag
}\left\vert r,n\right\rangle =C_{n+1}/C_{n}.$ Generally, the quantized
electric field (propagation direction $\overrightarrow{z}$) can be expressed
as $E\left( z,t\right) \varpropto \sin \left( kz\right) \left[ X\cos \left(
\omega t\right) +Y\sin \left( \omega t\right) \right] ,$ here the quadrature
operators $X$ and $Y$\ are associated with the amplitude and phase of field.
By virtue of the annihilation and creation operators, we have $X=\frac{%
a+a^{\dag }}{\sqrt{2}}$ and $Y=\frac{a-a^{\dag }}{i\sqrt{2}}.$\ Therefore,
the uncertainties in both quadratures are%
\begin{eqnarray}
\left\langle \Delta X\right\rangle ^{2} &=&\left\langle X^{2}\right\rangle
-\left\langle X\right\rangle ^{2}=M\left( n,r\right) +N\left( n,r\right) +%
\frac{1}{2},  \notag \\
\left\langle \Delta Y\right\rangle ^{2} &=&\left\langle Y^{2}\right\rangle
-\left\langle Y\right\rangle ^{2}=-M\left( n,r\right) +N\left( n,r\right) +%
\frac{1}{2},  \label{uncertainty} \\
\left\langle \Delta X\right\rangle \left\langle \Delta Y\right\rangle &=&%
\sqrt{\left\langle X^{2}\right\rangle -\left\langle X\right\rangle ^{2}}%
\times \sqrt{\left\langle Y^{2}\right\rangle -\left\langle Y\right\rangle
^{2}},  \notag
\end{eqnarray}%
where $N\left( n,r\right) =\frac{C_{n+1}}{C_{n}}-1.$ As is well known, the
coherent state is nearly a classical-like state and its quadratures are the
same minimum uncertainty, i.e. 
\begin{equation}
\left\langle \Delta X\right\rangle _{c}^{2}=\left\langle \Delta
Y\right\rangle _{c}^{2}=1/2,  \label{coherent}
\end{equation}%
which shows the expectation values of field contains only the noise of the
vacuum and this noise does not vanish. Reference to the vacuum nosie,
signal-to-noise ratio (SNR) $SNR\left( \hat{O}\right) =\log _{10}\left(
\left\langle \Delta \hat{O}\right\rangle ^{2}/\left\langle \Delta \hat{O}%
\right\rangle _{c}^{2}\right) $ is often used to measure the squeezing level
of a quantum field. From (\ref{uncertainty}) and (\ref{coherent}), we can
derive 
\begin{eqnarray}
SNR\left( X\right) &=&\log _{10}\left( \frac{\left\langle \Delta
X\right\rangle ^{2}}{\left\langle \Delta X\right\rangle _{c}^{2}}\right)
=\log _{10}\left[ 2M\left( n,r\right) +2N\left( n,r\right) +1\right] , 
\notag \\
SNR\left( Y\right) &=&\log _{10}\left( \frac{\left\langle \Delta
Y\right\rangle ^{2}}{\left\langle \Delta Y\right\rangle _{c}^{2}}\right)
=\log _{10}\left[ -2M\left( n,r\right) +2N\left( n,r\right) +1\right] ,
\label{snr} \\
SNR\left( X,Y\right) &=&\log _{10}\left( \frac{\left\langle \Delta
X\right\rangle \left\langle \Delta Y\right\rangle }{\left\langle \Delta
X\right\rangle _{c}\left\langle \Delta Y\right\rangle _{c}}\right) =\frac{1}{%
2}\log _{10}\left[ \left[ 2N\left( n,r\right) +1\right] ^{2}-4M^{2}\left(
n,r\right) \right] .  \notag
\end{eqnarray}

\begin{figure}[h]
\centering
\includegraphics[scale=0.50]{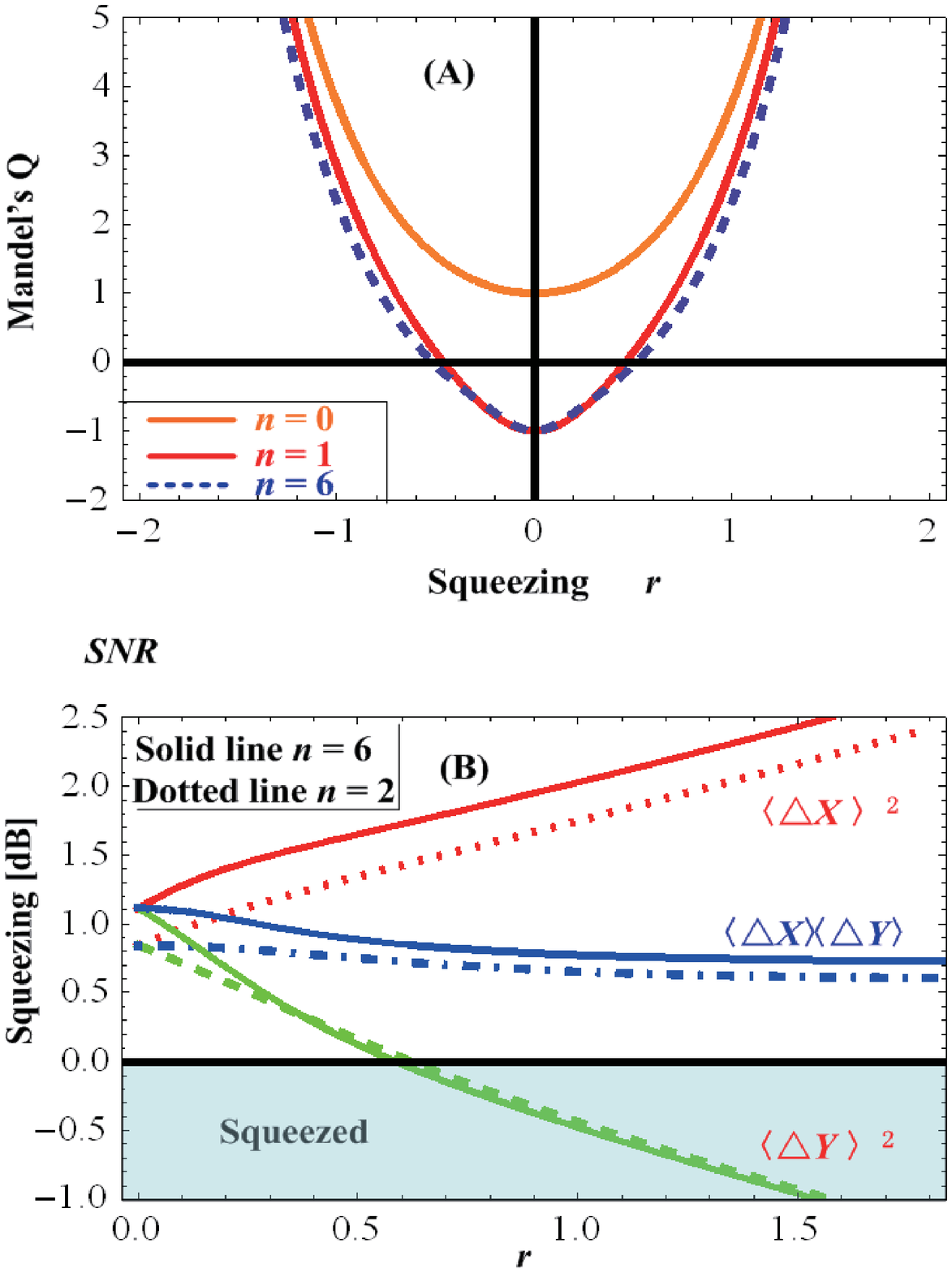}
\caption{\textsf{Nonclassical features of the ESVS. (A) The Mandel's $Q$
parameter with fixed values of $n=0,$ $1$ and $6$. (B) Quadrature
fluctuations $\left\langle \Delta X\right\rangle ^{2}$ (red line)$,$ $%
\left\langle \Delta Y\right\rangle ^{2}$ (green line) and $\left\langle
\Delta X\right\rangle \cdot \left\langle \Delta Y\right\rangle $ (blue line)
as the functions of the squeezing parameter $r$ with $n=2$ and $6$,
respectively.}}
\end{figure}

When $n=2$ and $6,$\ squeezing behaviours of the ESVS have been shown in
Figure 3(B). As a function of the squeezing parameter $r,$ $SNR$ is a
measure to the fluctuation level that is lower than the vacuum noise. Refer
to the vacuum field in units of dB, i.e. $SNR=0$ (see the black line in
Figure 3(B)), $SNR<0$\ means\ that the noise was squeezed in $Y$ quadrature.
With rising the value of $r,$\ $Y$ quadrature variance $\left\langle \Delta
Y\right\rangle ^{2}$ (green line) goes down gradually, as $X$ quadrature
variance (red line) goes up and joint variance $\left\langle \Delta
X\right\rangle \left\langle \Delta Y\right\rangle $ (blue line) reduces
slowly. When adding more and more photons, $SNR$ has a higher level. That
means squeezing behaviours of high-order ESVS can be dominated via squeezing
parameter $r.$

\section{Non-Gaussianity of the ESVS}

The criteria and estimation for the quantum non-Gaussianity have proposed by
Genoni in Refs.\cite{Genoni_2007_pra}. The non-Gaussianity can be evaluated
by%
\begin{equation}
\delta \left[ \rho \right] =\frac{D_{HS}^{2}\left[ \rho ,\tau \right] }{%
\text{Tr}\left[ \rho ^{2}\right] }=\frac{\text{Tr}\left[ \rho ^{2}\right] +%
\text{Tr}\left[ \tau ^{2}\right] -2\text{Tr}\left[ \rho \tau \right] }{2%
\text{Tr}\left[ \rho ^{2}\right] }=\frac{\mu \left[ \rho \right] +\mu \left[
\tau \right] -2\kappa \left[ \rho ,\tau \right] }{2\mu \left[ \rho \right] },
\label{ng}
\end{equation}%
in which $D_{HS}^{2}\left[ \rho ,\tau \right] $ is the Hilbert-Schmidt
distance between the estimated state $\rho $ and the reference Gaussian
state $\tau $, $\mu \left[ \rho \right] \equiv $Tr$\left[ \rho ^{2}\right] ,$
$\mu \left[ \tau \right] \equiv $Tr$\left[ \tau ^{2}\right] $, $\kappa \left[
\rho ,\tau \right] \equiv $Tr$\left[ \rho \tau \right] $. In general, the
reference Gaussian state is a displaced squeezed thermal state $\tau
=D\left( \epsilon \right) S\left( \varsigma \right) \nu \left( n_{t}\right)
S^{-1}\left( \varsigma \right) D^{-1}\left( \epsilon \right) $, where $%
D\left( \epsilon \right) =\exp \left( \epsilon a^{\dag }-\epsilon ^{\ast
}a\right) $ and $S\left( \varsigma \right) =\exp \left[ \frac{\varsigma }{2}%
\left( a^{\dag 2}-a^{2}\right) \right] $ are the displacement and squeezing
operators, repectively. $\nu \left( \bar{n}\right) =\left( 1+\bar{n}\right)
^{-1}\left[ \frac{\bar{n}}{1+\bar{n}}\right] ^{a^{\dag
}a}=\sum_{m=0}^{\infty }\frac{\bar{n}^{m}}{\left( 1+\bar{n}\right) ^{m+1}}%
\left\vert m\right\rangle \left\langle m\right\vert $ is the thermal state
with an $\bar{n}$ average number of photons. The parameters $\epsilon ,$ $%
\varsigma $ and $\bar{n}$\ are analytical functions of $n$ and $r.$ Firstly,
the reference Gaussian state $\tau $ will only work if 
\begin{equation}
X\left[ \rho \right] =X\left[ \tau \right] ,\text{ \ \ }\sigma \left[ \rho %
\right] =\sigma \left[ \tau \right] ,  \label{covariance}
\end{equation}%
in other words, their vector of mean values and the covariance matrix are
equal. In (\ref{covariance}), $X_{j}$ and $\sigma _{kj}$ denote the vector
of mean values and the covariance matrix of a quantum state, respectively,
satisfying%
\begin{equation}
X_{j}=\left\langle R_{j}\right\rangle ,\text{ \ \ }\sigma _{kj}=\left\langle
\left\{ R_{k},R_{j}\right\} \right\rangle -\left\langle R_{k}\right\rangle
\left\langle R_{j}\right\rangle ,  \label{vector}
\end{equation}%
where the real vector $R=\left( q_{1},p_{1},q_{2},p_{2},\cdots
,q_{n},p_{n}\right) ^{T}\ $has the commutation relations $\left[ R_{k},R_{j}%
\right] =i\Omega _{kj},$ symbol $\left\{ \cdots ,\cdots \right\} $ denotes
the anticommutator, $\left\langle \cdots \right\rangle $ is the expectation
value, and $\Omega _{kj}$ is the elements of the symplectic matrix $\Omega
=i\oplus _{k=1}^{n}\sigma _{2},$ $\sigma _{2}$ being the $y-$ Pauli matrix.
From the first equation of (\ref{covariance}), we have%
\begin{equation}
\text{Tr}\left[ a\rho \left( r,n\right) \right] =\text{Tr}\left[ a\tau
\left( \epsilon ,\varsigma ,\bar{n}\right) \right] ,\text{ \ \ Tr}\left[
a^{\dag }\rho \left( r,n\right) \right] =\text{Tr}\left[ a^{\dag }\tau
\left( \epsilon ,\varsigma ,\bar{n}\right) \right] ,  \label{tr1}
\end{equation}%
and considering%
\begin{equation}
D\left( \epsilon \right) aD^{-1}\left( \epsilon \right) =a-\epsilon ,\text{
\ \ \ }D\left( \epsilon \right) a^{\dag }D^{-1}\left( \epsilon \right)
=a^{\dag }-\epsilon ^{\ast },  \label{transform}
\end{equation}%
we obtain%
\begin{equation}
\text{Tr}\left[ a\tau \left( \epsilon ,\varsigma ,\bar{n}\right) \right] =%
\text{Tr}\left[ a\sum_{m=0}^{\infty }\frac{\bar{n}^{m}}{\left( 1+\bar{n}%
\right) ^{m+1}}D\left( \epsilon \right) S\left( \varsigma \right) \left\vert
m\right\rangle \left\langle m\right\vert S^{-1}\left( \varsigma \right)
D^{-1}\left( \epsilon \right) \right] =\epsilon =\text{Tr}\left[ a\rho
\left( r,n\right) \right] =0.  \label{displacement}
\end{equation}%
Indeed, the reference Gaussian state of the ESVS is a squeezed thermal state%
\begin{equation}
\tau \left( \varsigma ,\bar{n}\right) =S\left( \varsigma \right) \nu \left( 
\bar{n}\right) S^{-1}\left( \varsigma \right) =\sum_{m=0}^{\infty }\frac{%
\bar{n}^{m}}{\left( 1+\bar{n}\right) ^{m+1}}S\left( \varsigma \right)
\left\vert m\right\rangle \left\langle m\right\vert S^{-1}\left( \varsigma
\right) ,  \label{reference}
\end{equation}%
where the squeezing coefficient $\varsigma $\ and average photons number $%
\bar{n}$ shall be decided by the second equation of (\ref{covariance}), i.e. 
\begin{eqnarray}
\text{Tr}\left[ a^{2}\rho \left( r,n\right) \right] &=&\text{Tr}\left[
a^{2}\tau \left( \epsilon ,\varsigma ,\bar{n}\right) \right] ,\text{ \ \ } 
\notag \\
\text{Tr}\left[ a^{\dag 2}\rho \left( r,n\right) \right] &=&\text{Tr}\left[
a^{\dag 2}\tau \left( \epsilon ,\varsigma ,\bar{n}\right) \right] ,\ \ \ \ 
\label{tr2} \\
\text{Tr}\left[ a^{\dag }a\rho \left( r,n\right) \right] &=&\text{Tr}\left[
a^{\dag }a\tau \left( \epsilon ,\varsigma ,\bar{n}\right) \right] .  \notag
\end{eqnarray}%
Substituting Eqs.(\ref{reference}) and (\ref{do}) into Eq.(\ref{tr2}), we
have 
\begin{eqnarray}
\text{Tr}\left[ a^{2}\tau \left( \varsigma ,\bar{n}\right) \right] &=&\text{%
Tr}\left[ a^{\dag 2}\tau \left( \varsigma ,\bar{n}\right) \right] =\text{Tr}%
\left[ a^{2}\sum_{m=0}^{\infty }\frac{\bar{n}^{m}}{\left( 1+\bar{n}\right)
^{m+1}}S\left( \varsigma \right) \left\vert m\right\rangle \left\langle
m\right\vert S^{-1}\left( \varsigma \right) \right] =\left( 2\bar{n}%
+1\right) \sinh \varsigma \cosh \varsigma =M\left( n,r\right) ,
\label{parameter} \\
\text{Tr}\left[ a^{\dag }a\tau \left( \epsilon ,\varsigma ,\bar{n}\right) %
\right] &=&\bar{n}\cosh ^{2}\varsigma +\left( \bar{n}+1\right) \sinh
^{2}\varsigma =N(n,r),  \notag
\end{eqnarray}%
then it follows%
\begin{equation}
\varsigma =\frac{1}{4}\ln \left[ \frac{2N\left( n,r\right) +1+2M\left(
n,r\right) }{2N\left( n,r\right) +1-2M\left( n,r\right) }\right] ,\text{ \ \
\ \ }\bar{n}=\frac{\sqrt{\left[ 2N\left( n,r\right) +1\right]
^{2}-4M^{2}\left( n,r\right) }-1}{2},  \label{solution}
\end{equation}%
$M\left( n,r\right) $ and $N\left( n,r\right) $\ are also shown in (\ref{two}%
) and (\ref{uncertainty}).

Since the ESVS is a pure state, thus $\mu \left[ \rho \right] \equiv $Tr$%
\left[ \rho ^{2}\right] =$Tr$\left[ \rho \right] =1,$ while the squeezed
thermal states is mixture, $\mu \left[ \tau \right] =Tr\left[ \tau ^{2}%
\right] =\left( 1+2\bar{n}\right) ^{-1}.$ Substituting (\ref{reference}) and
(\ref{do}) into $\kappa \left[ \rho ,\tau \right] ,$ we can derive%
\begin{eqnarray}
\kappa \left[ \rho ,\tau \right] &=&C_{n}^{-1}\sum_{m=0}^{\infty }\frac{\bar{%
n}^{m}}{\left( 1+\bar{n}\right) ^{m+1}}\left\langle 0\right\vert
S^{-1}\left( r\right) a^{n}S\left( \varsigma \right) \left\vert
m\right\rangle \left\langle m\right\vert S^{-1}\left( \varsigma \right)
a^{\dag }{}^{n}S\left( r\right) \left\vert 0\right\rangle  \notag \\
&=&C_{n}^{-1}\sum_{m=0}^{\infty }\frac{\bar{n}^{m}}{\left( 1+\bar{n}\right)
^{m+1}}\left\vert \Lambda \left( r,\varsigma ,m,n\right) \right\vert ^{2},
\label{distance}
\end{eqnarray}%
where $\Lambda \left( r,\varsigma ,m,n\right) =\left\langle 0\right\vert
S^{-1}\left( r\right) a^{n}S\left( \varsigma \right) \left\vert
m\right\rangle \ $has an analytical expression, for details in Appendix C,%
\begin{eqnarray}
\Lambda \left( r,\varsigma ,m,n\right) &=&\frac{1}{\sqrt{m!\cosh \left(
\varsigma -r\right) }}\left( -i\sqrt{\frac{\sinh 2r}{4}}\right) ^{n}\left[ 1+%
\frac{\tanh \left( \varsigma -r\right) }{\tanh r}\right] ^{\frac{n}{2}}\left[
\frac{\tanh \left( \varsigma -r\right) }{2}\right] ^{\frac{m}{2}}  \notag \\
&&\times \sum_{k=0}^{m}\dbinom{m}{k}\frac{\left( 2i\right) ^{k}n!}{\left(
n-k\right) !}\left[ \frac{2}{\left( \tanh r+\tanh \left[ \varsigma -r\right]
\right) \sinh \left( 2\varsigma -2r\right) }\right] ^{\frac{k}{2}%
}H_{n-k}\left( 0\right) H_{m-k}\left( 0\right) .  \label{result}
\end{eqnarray}%
Finally, analytical expression of the non-Gaussianity of the ESVS can be
derived as%
\begin{equation}
\delta \left[ \rho \right] =\frac{1}{2}+\frac{1}{2+4\bar{n}}%
-C_{n}^{-1}\sum_{m=0}^{\infty }\frac{\bar{n}^{m}}{\left( 1+\bar{n}\right)
^{m+1}}\left\vert \Lambda \left( r,\varsigma ,m,n\right) \right\vert ^{2},
\label{final_ng}
\end{equation}%
which measures the deviation with reference to a Gaussian squeezed thermal
state. $\delta \left[ \rho \right] =0$ if and only if $\rho $ is a Gaussian
state. This point can be confirmed in Figure 4(A) with $n=0$ (pink line). In
Figure 4(A), from the bottom to the top we can see that non-Gaussianity of
the ESVS can be enhanced by adding more and more photon to squeezed vacuum
field. Besides added photon number, non-Gaussianity also strongly depends on
the squeezing level. When $\left\vert r\right\vert \rightarrow 0,$
non-Gaussianity of the ESVS has a higher performance. Similar results can be
found for the case of odd photon-addition (see Figure 4(B)).

\begin{figure}[h]
\centering
\includegraphics[scale=0.50]{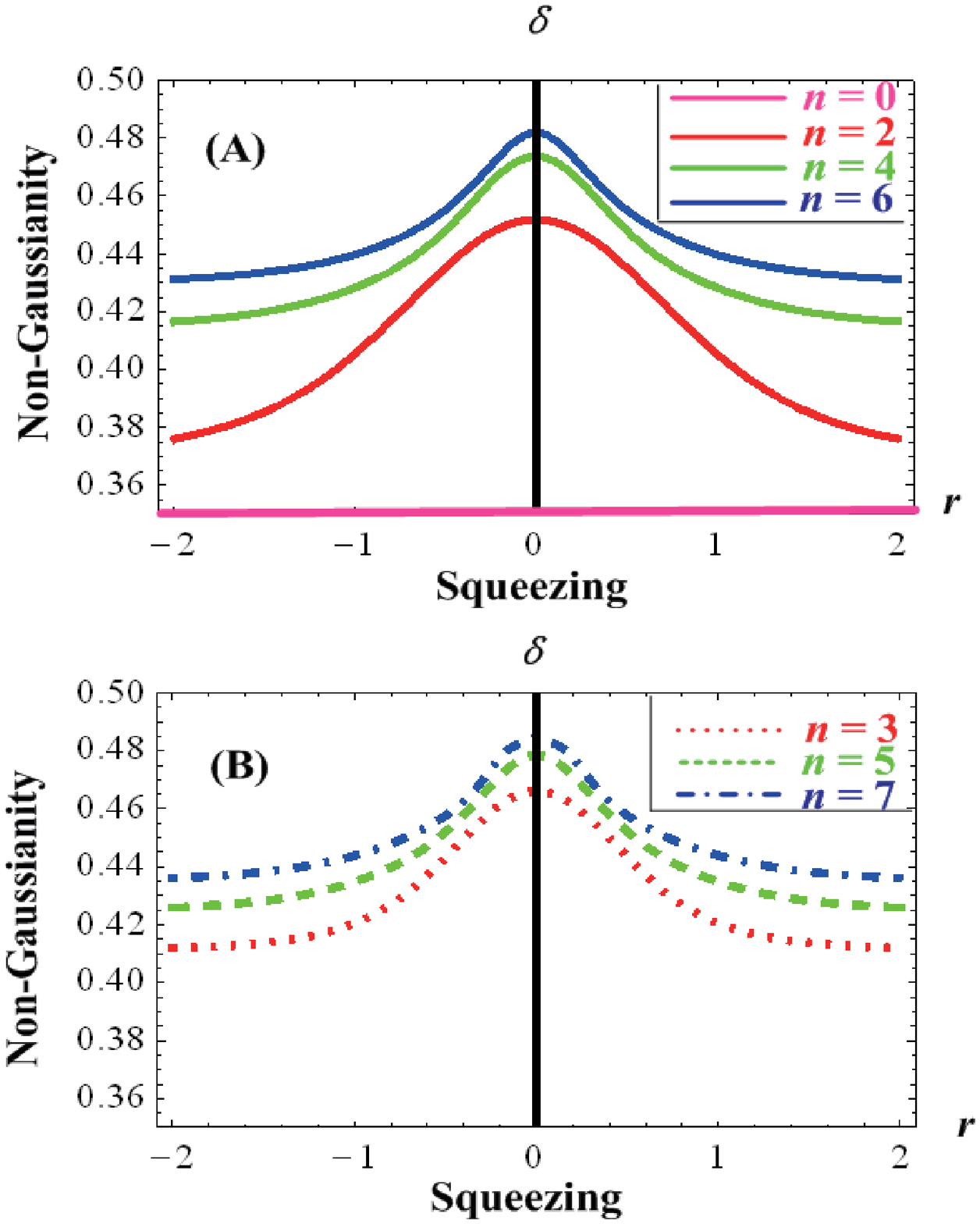}
\caption{\textsf{Non-Gaussianity of the ESVS with fixed $n$ and the varying
squeezing parameter $r\in \lbrack -2,2].$ (A): Adding even photon $n=0,$ $2,$
$4,$ $6\ $correspond to the solid lines from the buttom to the top. The
lowest line means that non-Gaussianity of a Gaussian state is always zero
(pink line). (B): Adding odd photon $n=3,$ $5,$ $7\ $correspond to the
dotted line, dashed line and dot-dashed line, respectively.}}
\end{figure}

\section{Fidelity Between the ESVS and the Photon-subtraction Squeezed
Vacuum State (PSSVS)}

Eq.(\ref{sq}) indicates that the squeezed field is an approximation of all
even number photon superposition states. If adding one single photon to this
squeezed field, we can simplify $a^{\dag }S\left( r\right) \left\vert
0\right\rangle \rightarrow \alpha _{1}\left\vert 1\right\rangle +\beta
_{1}\left\vert 3\right\rangle +\gamma _{1}\left\vert 5\right\rangle +\cdots
, $ which is a superposition of odd number photon states shown in Eq.(\ref%
{csq}). When subtracting one single photon from the squeezed field, we can
see 
\begin{equation}
aS\left( r\right) \left\vert 0\right\rangle \rightarrow \sum_{n=0}\frac{%
\sqrt{\left( 2n-1\right) !}}{2^{n-1}\left( n-1\right) !}\tanh
^{n}r\left\vert 2n-1\right\rangle \rightarrow \alpha _{2}\left\vert
1\right\rangle +\beta _{2}\left\vert 3\right\rangle +\gamma _{2}\left\vert
5\right\rangle +\cdots ,  \label{spssvs}
\end{equation}%
where $\alpha _{j},$ $\beta _{j}$ and $\gamma _{j}\ \left( j=1,2\right) $
are real. Therefore, from the point of view of the photon number
distribution, the ESVS is close to the PSSVS. The similarity of these states
can be estimated by difining the fidelity $F,$ i.e. 
\begin{equation}
F=\frac{Tr\left[ \rho _{0}\rho \left( r,n\right) \right] }{Tr\left[ \rho
_{0}^{2}\right] },  \label{fidelity}
\end{equation}%
where $\rho \left( r,n\right) $ is the ESVS shown in Eq.(\ref{do}), $\rho
_{0}$ denotes the PSSVS, 
\begin{equation}
\rho _{0}=C_{m}^{-1}a^{m}S\left( \lambda \right) \left\vert 0\right\rangle
\left\langle 0\right\vert S^{-1}\left( \lambda \right) a^{\dag }{}^{m},
\label{pssvs}
\end{equation}%
$C_{m}=m!\left( i\sinh \lambda \right) ^{m}P_{m}\left( -i\sinh \lambda
\right) $ is the normalized constant. Noting $Tr\left[ \rho _{0}^{2}\right]
=Tr\left[ \rho _{0}\right] =1,$ we have%
\begin{eqnarray}
F &=&Tr\left[ \rho _{0}\rho \left( r,n\right) \right] =Tr\left[
C_{n}^{-1}C_{m}^{-1}a^{m}S\left( \lambda \right) \left\vert 0\right\rangle
\left\langle 0\right\vert S^{-1}\left( \lambda \right) a^{\dag
}{}^{m}a^{\dag }{}^{n}S\left( r\right) \left\vert 0\right\rangle
\left\langle 0\right\vert S^{-1}\left( r\right) a^{n}\right]  \notag \\
&=&C_{n}^{-1}C_{m}^{-1}\left\langle 0\right\vert S^{-1}\left( r\right)
a^{m+n}S\left( \lambda \right) \left\vert 0\right\rangle \left\langle
0\right\vert S^{-1}\left( \lambda \right) a^{\dag }{}^{m+n}S\left( r\right)
\left\vert 0\right\rangle  \label{inner} \\
&\equiv &C_{n}^{-1}C_{m}^{-1}\left\vert \Gamma \left( r,\lambda ,m,n\right)
\right\vert ^{2}  \notag
\end{eqnarray}%
where $\Gamma \left( r,\lambda ,m,n\right) =\left\langle 0\right\vert
S^{-1}\left( r\right) a^{m+n}S\left( \lambda \right) \left\vert
0\right\rangle .$ Comparing $\Gamma \left( r,\lambda ,m,n\right) $ and $%
\Lambda \left( r,\varsigma ,m,n\right) =\left\langle 0\right\vert
S^{-1}\left( r\right) a^{n}S\left( \varsigma \right) \left\vert
m\right\rangle $ shown in Eq.(\ref{distance}), we can obtain the analytical
expression of $\Gamma \left( r,\lambda ,m,n\right) $ via the replacement of $%
\varsigma \rightarrow \lambda ,$ $m\rightarrow 0,$ and $n\rightarrow m+n$ in
Eq.(\ref{result}), i.e. 
\begin{equation}
\Gamma \left( r,\lambda ,m,n\right) =\frac{1}{\sqrt{\cosh \left( \lambda
-r\right) }}\left( -i\sqrt{\frac{\sinh 2\lambda }{4}}\right) ^{m+n}\left[ 1+%
\frac{\tanh \left( \lambda -r\right) }{\tanh r}\right] ^{\frac{m+n}{2}%
}H_{m+n}\left[ 0\right] .  \label{tao}
\end{equation}%
As a result, the fidelity can be derived as%
\begin{equation}
F=C_{n}^{-1}C_{m}^{-1}\frac{1}{\cosh \left( \lambda -r\right) }\left( \frac{%
\sinh 2r}{4}\right) ^{m+n}\left[ 1+\frac{\tanh \left( \lambda -r\right) }{%
\tanh r}\right] ^{m+n}\left[ H_{m+n}\left( 0\right) \right] ^{2},  \label{ff}
\end{equation}%
which is an analytical expression for the fidelity between the ESVS and the
PSSVS with variance of parameters $n$, $m$, $\lambda $ and $r.$ On the left
column of Figure 5 (see Figure 5(A)---5(C)), the values of photon-addition
number $n$ and photon-subtraction number $m$ are independent. In Figure
5(A), taking $n=m=2,$ the fidelity as a function of the squeezing parameter $%
r$ can be obtianed by taking different values of $\lambda =0.2,$ $0.5,$ $1.0$
and $1.5,$ respectively. When $\lambda =1.5,$ the optimal fidelity is $%
0.992613$ at $r=1.4758.$ If taking $n=2,$ $m=4,$ the fidelity has been shown
in Figure 5(B), and under the same conditions ($\lambda =1.5$) the optimal
fidelity is $0.971793$ at $r=1.76518.$ Figure 5(C) describes the case of $%
n=2,$ $m=6.$

\begin{figure}[h]
\centering
\includegraphics[scale=0.50]{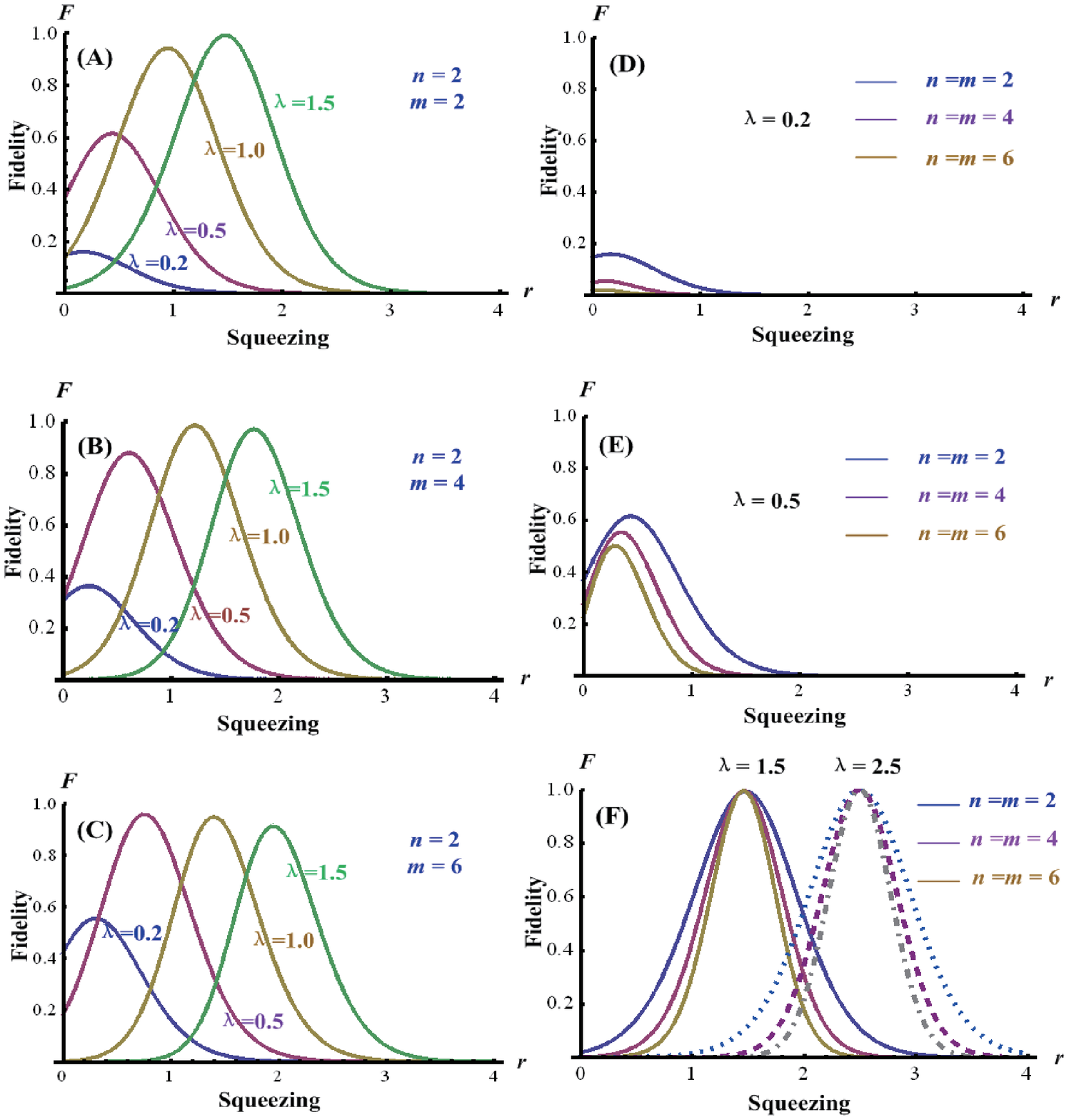}
\caption{\textsf{Fidelity between the ESVS and PSSVS with variance of
parameters $n$, $m$, $\protect\lambda $ and $r.$ With fixed values of $%
\protect\lambda =0.2,$ $0.5,$ $1.0$ and $1.5,$ the left column of Figure 5
exhibits the continuously varying fidelity with (A) $n=m=2$, (B) $n=2,$ $m=4$%
, (C) $n=2,$ $m=6$. For the case of the same values of $n=m=2,$ $4$ and $6,$
the right column exhibits the fidelity with different squeezing (D) $\protect%
\lambda =0.2$, (E) $\protect\lambda =0.4$, (F) $\protect\lambda =1.5$ and $%
2.5$.}}
\end{figure}

If always taking the same values of $n$ and $m,$\ the fidelity is a
continuously varying with the squeezing $r$ and $\lambda $ (with the
different values of $\lambda =0.2,$ $0.5,$ $1.5$ and $2.5).$ The details can
be found in Figure 5(D)--5(E). With the increase of $\lambda ,$ the optimal
fidelity for $r$ will be pretty close to $1$. For example of $n=m=2$ and $%
\lambda =2.5,$ the optimal fidelity is $0.99987$ at $r=2.49645.$ When $n=4,$ 
$m=6,$ and taking the different values of $\lambda $ and $r$, the fidelity
distribution has been shown in Figure 6(A). Figure 6(B) depicts the contour
of fidelity with respect to $\lambda $ and $r.$

\begin{figure}[h]
\centering
\includegraphics[scale=0.50]{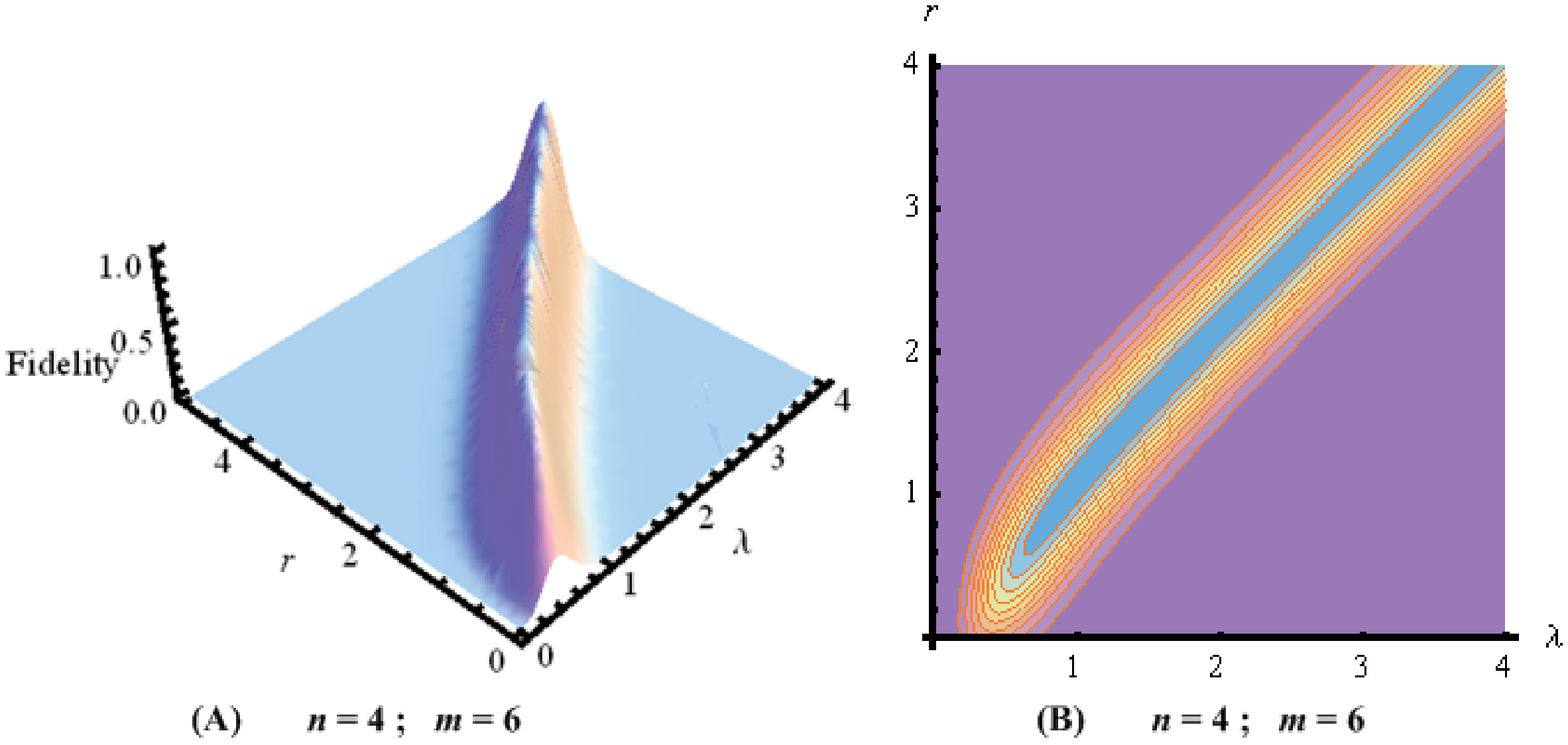}
\caption{\textsf{Fidelity distribution (A) and its contour (B) with $n=4$
and $m=6.$ As a function of the squeezing parameters $\protect\lambda $ and $%
r,$ the optimal fidelity can be decided by adjusting parameters for
different cases. In Figure 6(A), the ridge is the descrption of optimal
fidelity. In Figure 6(B), the smaller blue region means much higher fidelity.%
}}
\end{figure}

\section{Conclusions and Remarks}

Non-Gaussian operations and non-Gaussian quantum states have been proved to
be very effective means and useful resources in continuous variable (CV)
quantum information processes (QIP). In this work, we have introduced the
excited squeezed vacuum state (ESVS), which can be ustilized to describe
quantum light field emitted from multiphoton quantum process occurred in
some restricted quantum systems. And then we have exhibited its nonclassical
properities (e.g. the phase space Wigner distribution, photo number
distributions, the second-order autocorrelation Mandel's $Q$ parameter, the
quadrature fluctuations). We have also investigated how to quantify the
non-Gaussianity of the ESVS. After all, with enhanced non-Gaussian
properties may constitute powerful resources for many quantum information
tasks \cite%
{Cerf_2005_prl,Ourjoumtsev_2007_nature,Dong_2008_nature,Lund_2008_prl,Xiang_2010_np}%
. Due to the similar photon number distribution, we have examined the
fidelity between the ESVS and the PSSVS, and obtianed an analytical
expression of fidelity, from which the optimal fidelity can be decided by
monitoring the relevant parameters. With current technology of experimental
study, high fidelity preparation of the ESVS can be replaced by the PSSVS
generation. In fact, the photon-subtraction operation is relatively easy to
implement via the simple optical devices (e.g. beamsplitter) and coincidence
measurement.

\section{Acknowledgements}

This work has been supported in part by the Natural Science Foundation of
China (No. 61203061, 11204004, 61374091, 61403362, and 61473199),
Outstanding Young Talent Foundation of Anhui Province Colleges and
Universities No.2012SQRL040, and also Natural Science Foundation of Anhui
Province Colleges \& Universities under grant KJ2012Z035. The authors
acknowledge fruitful discussions about physics meaning with Prof. Guoyong
Xiang, Prof. Chaoyang Lu and their group members.

\section{Appendix A: Derivation of Eq.(\protect\ref{wigner})}

Noting the following transformations%
\begin{eqnarray}
S^{-1}\left( r\right) aS\left( r\right) &=&a\cosh r+a^{\dag }\sinh r,  \notag
\\
S^{-1}\left( r\right) a^{\dag }S\left( r\right) &=&a^{\dag }\cosh r+a\sinh r,
\label{aa1}
\end{eqnarray}%
we have%
\begin{eqnarray}
\rho \left( \lambda ,n\right) &=&C_{n}^{-1}a^{\dag }{}^{n}S\left( \lambda
\right) \left\vert 0\right\rangle \left\langle 0\right\vert S^{-1}\left(
\lambda \right) a^{n}  \notag \\
&=&C_{n}^{-1}S\left( \lambda \right) \left( a^{\dag }\cosh r+a\sinh r\right)
^{n}\left\vert 0\right\rangle \left\langle 0\right\vert \left( a\cosh
r+a^{\dag }\sinh r\right) ^{n}S^{-1}\left( \lambda \right) .  \label{aa2}
\end{eqnarray}%
Thus the Wigner function of $\rho \left( \lambda ,n\right) $ can be written
as%
\begin{eqnarray}
W\left( \alpha ,\alpha ^{\ast }\right) &=&Tr\left[ \rho \left( \lambda
,n\right) \Delta \left( \alpha ,\alpha ^{\ast }\right) \right]
=C_{n}^{-1}\left\langle 0\right\vert S^{-1}\left( \lambda \right)
a^{n}\Delta \left( \alpha ,\alpha ^{\ast }\right) a^{\dag }{}^{n}S\left(
\lambda \right) \left\vert 0\right\rangle  \notag \\
&=&C_{n}^{-1}\left\langle 0\right\vert \left( a\cosh r+a^{\dag }\sinh
r\right) ^{n}S^{-1}\left( \lambda \right) \Delta \left( \alpha ,\alpha
^{\ast }\right) S\left( \lambda \right) \left( a^{\dag }\cosh r+a\sinh
r\right) ^{n}\left\vert 0\right\rangle .  \label{aa3}
\end{eqnarray}%
Considering (\ref{7}) and directly using the transform in (\ref{invariance}%
), we have%
\begin{eqnarray}
S^{-1}\left( \lambda \right) \Delta \left( \alpha ,\alpha ^{\ast }\right)
S\left( \lambda \right) &=&\frac{1}{2}S^{-1}\left( \lambda \right) 
\begin{array}{c}
\dot{\vdots}%
\end{array}%
\delta \left( a-\alpha \right) \delta \left( a^{\dagger }-\alpha ^{\ast
}\right) 
\begin{array}{c}
\dot{\vdots}%
\end{array}%
S\left( \lambda \right)  \notag \\
&=&\frac{1}{2}%
\begin{array}{c}
\dot{\vdots}%
\end{array}%
\delta \left( a\cosh r+a^{\dag }\sinh r-\alpha \right) \delta \left( a^{\dag
}\cosh r+a\sinh r-\alpha ^{\ast }\right) 
\begin{array}{c}
\dot{\vdots}%
\end{array}
\label{aa4} \\
&=&\frac{1}{2}%
\begin{array}{c}
\dot{\vdots}%
\end{array}%
\delta \left( a-\alpha \cosh r+\alpha ^{\ast }\sinh r\right) \delta \left(
a^{\dag }-\alpha ^{\ast }\cosh r+\alpha \sinh r\right) 
\begin{array}{c}
\dot{\vdots}%
\end{array}%
.  \notag
\end{eqnarray}%
The first equation of Eq.(\ref{7}) indicates%
\begin{equation}
S^{-1}\left( \lambda \right) \Delta \left( \alpha ,\alpha ^{\ast }\right)
S\left( \lambda \right) \rightarrow \Delta \left( z,z^{\ast }\right) =\int 
\frac{d^{2}\beta }{\pi ^{2}}\left\vert z+\beta \right\rangle \left\langle
z-\beta \right\vert e^{z\beta ^{\ast }-z^{\ast }\beta },  \label{aa5}
\end{equation}%
where $z=\alpha \cosh r-\alpha ^{\ast }\sinh r.$ Thus we have%
\begin{equation}
W\left( \alpha ,\alpha ^{\ast }\right) =C_{n}^{-1}\left\langle 0\right\vert
\left( a\cosh r+a^{\dag }\sinh r\right) ^{n}\int \frac{d^{2}\beta }{\pi ^{2}}%
\left\vert z+\beta \right\rangle \left\langle z-\beta \right\vert e^{z\beta
^{\ast }-z^{\ast }\beta }\left( a^{\dag }\cosh r+a\sinh r\right)
^{n}\left\vert 0\right\rangle .  \label{aa7}
\end{equation}%
Ref. \cite{Meng_2007_pla} has given a formular 
\begin{equation}
\left( fa+ga^{\dag }\right) ^{n}=\left( -i\sqrt{\frac{fg}{2}}\right)
^{n}\colon H_{n}\left( i\sqrt{\frac{f}{2g}}a+i\sqrt{\frac{g}{2f}}a^{\dag
}\right) \colon ,  \label{aa6}
\end{equation}%
then it follows%
\begin{eqnarray}
\left( a^{\dag }\cosh r+a\sinh r\right) ^{n} &=&\left( -i\sqrt{\frac{\sinh 2r%
}{4}}\right) ^{n}\colon H_{n}\left( i\sqrt{\frac{\sinh r}{2\cosh r}}a+i\sqrt{%
\frac{\cosh r}{2\sinh r}}a^{\dag }\right) \colon ,  \notag \\
\left( a\cosh r+a^{\dag }\sinh r\right) ^{n} &=&\left( -i\sqrt{\frac{\sinh 2r%
}{4}}\right) ^{n}\colon H_{n}\left( i\sqrt{\frac{\cosh r}{2\sinh r}}a+i\sqrt{%
\frac{\sinh r}{2\cosh r}}a^{\dag }\right) \colon .  \label{aa8}
\end{eqnarray}%
Substituting (\ref{aa8}) into (\ref{aa7}), we can see%
\begin{eqnarray}
W\left( \alpha ,\alpha ^{\ast }\right) &=&C_{n}^{-1}\left( -\frac{\sinh 2r}{4%
}\right) ^{n}\int \frac{d^{2}\beta }{\pi ^{2}}H_{n}\left[ i\sqrt{\frac{\cosh
r}{2\sinh r}}\left( z+\beta \right) \right] H_{n}\left[ i\sqrt{\frac{\cosh r%
}{2\sinh r}}\left( z^{\ast }-\beta ^{\ast }\right) \right]  \label{aa9} \\
&&\times \exp \left[ -\left\vert z\right\vert ^{2}-\left\vert \beta
\right\vert ^{2}+z\beta ^{\ast }-z^{\ast }\beta \right] .  \notag
\end{eqnarray}%
Recall the generating function of the Hermite polynomials%
\begin{equation}
H_{n}\left( x\right) =\frac{\partial {}^{n}}{\partial t{}^{n}}\left. \exp
\left( 2xt-t^{2}\right) \right\vert _{t=0},  \label{aa10}
\end{equation}%
we have%
\begin{eqnarray}
W\left( \alpha ,\alpha ^{\ast }\right) &=&C_{n}^{-1}\left( -\frac{\sinh 2r}{4%
}\right) ^{n}\frac{\partial {}^{2n}}{\partial t{}^{n}\partial \tau {}^{n}}%
\int \frac{d^{2}\beta }{\pi ^{2}}\exp \left[ -\left\vert z\right\vert
^{2}-\left\vert \beta \right\vert ^{2}+z\beta ^{\ast }-z^{\ast }\beta \right]
\notag \\
&&\times \left. \exp \left[ 2i\sqrt{\frac{\cosh r}{2\sinh r}}\left( z+\beta
\right) t-t^{2}+2i\sqrt{\frac{\cosh r}{2\sinh r}}\left( z^{\ast }-\beta
^{\ast }\right) \tau -\tau ^{2}\right] \right\vert _{t,\tau =0},
\label{aa11}
\end{eqnarray}%
and noting the following integral formula%
\begin{equation}
\int \frac{d^{2}\alpha }{\pi }\exp \left[ h\left\vert \alpha \right\vert
^{2}+s\alpha +\eta \alpha ^{\ast }+f\alpha ^{2}+g\alpha ^{\ast 2}\right] =%
\frac{1}{\sqrt{h^{2}-4fg}}\exp \left[ \frac{-hs\eta +s^{2}g+\eta ^{2}f}{%
h^{2}-4fg}\right] ,  \label{aa12}
\end{equation}%
whose convergent condition is Re$\left[ h\pm f\pm g\right] <0$ and Re$\left[ 
\frac{h^{2}-4fg}{h\pm f\pm g}\right] <0,$ we can derive%
\begin{eqnarray}
W\left( \alpha ,\alpha ^{\ast }\right) &=&C_{n}^{-1}\left( -\frac{\sinh 2r}{4%
}\right) ^{n}\exp \left[ -2\left\vert z\right\vert ^{2}\right]  \notag \\
&&\times \left. \frac{\partial {}^{2n}}{\partial t{}^{n}\partial \tau {}^{n}}%
\exp \left[ -t^{2}-\tau ^{2}+4i\sqrt{\frac{\cosh r}{2\sinh r}}zt+4i\sqrt{%
\frac{\cosh r}{2\sinh r}}z^{\ast }\tau +\frac{2\cosh r}{\sinh r}t\tau \right]
\right\vert _{t,\tau =0}.  \label{aa13}
\end{eqnarray}%
Using Eq.(\ref{aa10}) again, we have%
\begin{eqnarray}
W\left( \alpha ,\alpha ^{\ast }\right) &=&C_{n}^{-1}\left( -\frac{\sinh 2r}{4%
}\right) ^{n}\exp \left[ -2\left\vert z\right\vert ^{2}\right]  \label{aa14}
\\
&&\times \left. \frac{\partial {}^{n}}{\partial t{}^{n}}\left\{ \left[
H_{n}\left( 2i\sqrt{\frac{\cosh r}{2\sinh r}}z^{\ast }+\frac{\cosh r}{\sinh r%
}t\right) \right] \exp \left[ -t^{2}+4i\sqrt{\frac{\cosh r}{2\sinh r}}zt%
\right] \right\} \right\vert _{t=0},  \notag
\end{eqnarray}%
then it follows%
\begin{eqnarray}
W\left( \alpha ,\alpha ^{\ast }\right) &=&C_{n}^{-1}\left( -\frac{\sinh 2r}{4%
}\right) ^{n}\exp \left[ -2\left\vert z\right\vert ^{2}\right] \sum_{m=0}^{n}%
\dbinom{n}{m}  \notag \\
&&\times \left. \frac{\partial ^{m}}{\partial t^{m}}H_{n}\left( i\sqrt{\frac{%
2}{\tanh r}}z^{\ast }+\frac{t}{\tanh r}\right) \times \frac{\partial ^{n-m}}{%
\partial t^{n-m}}\exp \left[ -t^{2}+2i\sqrt{\frac{2}{\tanh r}}zt\right]
\right\vert _{t=0}.  \label{aa15}
\end{eqnarray}%
By virtue of the recurrence relation of $H_{n}\left( x\right) $%
\begin{equation}
\frac{d^{l}}{dx^{l}}H_{n}\left( x\right) =\frac{2^{l}n!}{\left( n-l\right) !}%
H_{n-l}\left( x\right) ,  \label{aa16}
\end{equation}%
we have%
\begin{equation}
W\left( \alpha ,\alpha ^{\ast }\right) =C_{n}^{-1}\left( \frac{\sinh 2r}{4}%
\right) ^{n}\exp \left( -2\left\vert z\right\vert ^{2}\right) \sum_{m=0}^{n}%
\dbinom{n}{m}\frac{2^{m}n!}{\left( n-m\right) !}\left( -\coth r\right)
^{m}\left\vert H_{n-m}\left( -i\sqrt{\frac{2}{\tanh r}}z\right) \right\vert
^{2}  \label{aa17}
\end{equation}%
and substituting $z=\alpha \cosh r-\alpha ^{\ast }\sinh r$ into the above
equation, finally we obtain the Wigner function of the ESVS%
\begin{eqnarray}
W\left( \alpha ,\alpha ^{\ast }\right) &=&\frac{1}{P_{n}\left( \cosh
r\right) }\left( \frac{\sinh r}{2}\right) ^{n}\exp \left[ -2\left\vert
\alpha \cosh r-\alpha ^{\ast }\sinh r\right\vert ^{2}\right]  \notag \\
&&\times \sum_{m=0}^{n}\dbinom{n}{m}\frac{2^{m}n!}{\left( n-m\right) !}%
\left( -\coth r\right) ^{m}\left\vert H_{n-m}\left[ -i\sqrt{\frac{2}{\tanh r}%
}\left( \alpha \cosh r-\alpha ^{\ast }\sinh r\right) \right] \right\vert
^{2}.  \label{aa18}
\end{eqnarray}

\section{Appendix B: Derivation of Expected Value in Eq.(\protect\ref%
{general})}

From (\ref{aa6}), we have 
\begin{eqnarray}
\left( a^{\dag }\cosh r+a\sinh r\right) ^{n+l} &=&\left( -i\sqrt{\frac{\sinh
r\cosh r}{2}}\right) ^{n+l}\colon H_{n+l}\left( i\sqrt{\frac{\sinh r}{2\cosh
r}}a+i\sqrt{\frac{\cosh r}{2\sinh r}}a^{\dag }\right) \colon ,  \notag \\
\left( a\cosh r+a^{\dag }\sinh r\right) ^{n+k} &=&\left( -i\sqrt{\frac{\sinh
r\cosh r}{2}}\right) ^{n+k}\colon H_{n+k}\left( i\sqrt{\frac{\cosh r}{2\sinh
r}}a+i\sqrt{\frac{\sinh r}{2\cosh r}}a^{\dag }\right) \colon .  \label{ab1}
\end{eqnarray}%
Inserting the completeness $\pi ^{-1}\int d^{2}\beta \left\vert \beta
\right\rangle \left\langle \beta \right\vert =1$ of coherent state $%
\left\vert \beta \right\rangle $ into (\ref{9}), we have%
\begin{eqnarray}
\left\langle \lambda ,n\right\vert a^{k}a^{\dag }{}^{l}\left\vert \lambda
,n\right\rangle &=&C_{n}^{-1}\left( -i\sqrt{\frac{\sinh 2r}{4}}\right)
^{k}\left( -i\sqrt{\frac{\sinh 2r}{4}}\right) ^{l}\left( -\frac{\sinh 2r}{4}%
\right) ^{n}  \notag \\
&&\times \int \frac{d^{2}\beta }{\pi ^{2}}H_{n+k}\left( i\sqrt{\frac{\cosh r%
}{2\sinh r}}\beta \right) H_{n+l}\left( i\sqrt{\frac{\cosh r}{2\sinh r}}%
\beta ^{\ast }\right) \exp \left( -\left\vert \beta \right\vert ^{2}\right) .
\label{ab2}
\end{eqnarray}%
Using Eq.(\ref{aa10}) twice, we have%
\begin{eqnarray}
\left\langle \lambda ,n\right\vert a^{k}a^{\dag }{}^{l}\left\vert \lambda
,n\right\rangle &=&C_{n}^{-1}\left( -i\sqrt{\frac{\sinh 2r}{4}}\right)
^{k}\left( -i\sqrt{\frac{\sinh 2r}{4}}\right) ^{l}\left( -\frac{\sinh 2r}{4}%
\right) ^{n}\frac{\partial {}^{n+k}}{\partial t{}^{n+k}}\frac{\partial
{}^{n+l}}{\partial \tau {}^{n+l}}  \notag \\
&&\times \left. \int \frac{d^{2}\beta }{\pi ^{2}}\exp \left[ -\left\vert
\beta \right\vert ^{2}-t^{2}+2i\sqrt{\frac{\cosh r}{2\sinh r}}\beta t-\tau
^{2}+2i\sqrt{\frac{\cosh r}{2\sinh r}}\beta ^{\ast }\tau \right] \right\vert
_{t,\tau =0},  \label{ab3}
\end{eqnarray}%
and using the following integral formula%
\begin{equation*}
\int \frac{d^{2}\alpha }{\pi }\exp \left[ h\left\vert \alpha \right\vert
^{2}+s\alpha +\eta \alpha ^{\ast }\right] =\frac{1}{h}\exp \left[ -\frac{%
s\eta }{h}\right] ,\text{ \ Re}\left[ h\right] <0,
\end{equation*}%
we can obtain%
\begin{eqnarray}
\left\langle \lambda ,n\right\vert a^{k}a^{\dag }{}^{l}\left\vert \lambda
,n\right\rangle &=&C_{n}^{-1}\left( -i\sqrt{\frac{\sinh 2r}{4}}\right)
^{k}\left( -i\sqrt{\frac{\sinh 2r}{4}}\right) ^{l}\left( -\frac{\sinh 2r}{4}%
\right) ^{n}\left. \frac{\partial {}^{n+k}}{\partial t{}^{n+k}}\frac{%
\partial {}^{n+l}}{\partial \tau {}^{n+l}}\exp \left[ -t^{2}-\tau ^{2}-2%
\frac{\cosh r}{\sinh r}t\tau \right] \right\vert _{t,\tau =0}  \notag \\
&=&C_{n}^{-1}\left( -i\sqrt{\frac{\sinh 2r}{4}}\right) ^{k}\left( i\sqrt{%
\frac{\sinh 2r}{4}}\right) ^{l}\left( \frac{\sinh 2r}{4}\right) ^{n}\left. 
\frac{\partial {}^{n+k}}{\partial t{}^{n+k}}\left[ H_{n+l}\left( \frac{\cosh
r}{\sinh r}t\right) \exp \left[ -t^{2}\right] \right] \right\vert _{t=0} 
\notag \\
&=&C_{n}^{-1}\left( -i\sqrt{\frac{\sinh 2r}{4}}\right) ^{k}\left( i\sqrt{%
\frac{\sinh 2r}{4}}\right) ^{l}\left( \frac{\sinh 2r}{4}\right) ^{n}
\label{ab4} \\
&&\times \sum_{m=0}^{n+k}\dbinom{n+k}{m}\coth ^{m}r\times \left[ \frac{%
\partial ^{m}}{\partial t^{m}}H_{n+l}\left( t\right) \right] \times \left. 
\frac{\partial ^{n+k-m}}{\partial t^{n+k-m}}\exp \left( -t^{2}\right)
\right\vert _{t=0}.  \notag
\end{eqnarray}%
Then using the recurrence relation in (\ref{aa16}) and the differential form
of the Hermite polynomials in Eq. (\ref{5}), we can derive%
\begin{eqnarray}
\left\langle \lambda ,n\right\vert a^{k}a^{\dag }{}^{l}\left\vert \lambda
,n\right\rangle &=&C_{n}^{-1}\left( -i\sqrt{\frac{\sinh 2r}{4}}\right)
^{k}\left( i\sqrt{\frac{\sinh 2r}{4}}\right) ^{l}\left( \frac{\sinh 2r}{4}%
\right) ^{n}  \notag \\
&&\times \sum_{m=0}^{n+k}\dbinom{n+k}{m}\frac{\left( -1\right)
^{n+k-m}2^{m}\left( n+l\right) !\coth ^{m}r}{\left( n+l-m\right) !}%
H_{n+l-m}\left( 0\right) H_{n+k-m}\left( 0\right) .  \label{ab5}
\end{eqnarray}

\section{Appendix C: Derivation of Non-Gaussianity in Eq.(\protect\ref%
{result})}

Noting Eqs.(\ref{aa1}), (\ref{aa6}) and (\ref{aa8}), we have%
\begin{eqnarray}
\Lambda \left( r,\varsigma ,m,n\right) &\equiv &\left\langle 0\right\vert
S^{-1}\left( r\right) a^{n}S\left( \varsigma \right) \left\vert
m\right\rangle =\left\langle 0\right\vert \left( a\cosh r+a^{\dag }\sinh
r\right) ^{n}S\left( \varsigma -r\right) \left\vert m\right\rangle  \notag \\
&=&\left( -i\sqrt{\frac{\sinh r\cosh r}{2}}\right) ^{n}\left\langle
0\right\vert H_{n}\left( i\sqrt{\frac{\cosh r}{2\sinh r}}a\right) S\left(
\nu \right) \left\vert m\right\rangle ,  \label{ac1}
\end{eqnarray}%
where $\nu =\varsigma -r.$ Inserting the completeness relation of coherent
state into the above equation and using $\left\vert m\right\rangle =\left. 
\frac{1}{\sqrt{m!}}\frac{d^{m}}{d\alpha ^{m}}\left\vert \alpha \right\rangle
\right\vert _{\alpha =0}$ ($\left\vert \alpha \right\rangle =\exp \left(
\alpha a^{\dag }\right) \left\vert 0\right\rangle $)$,$ we have%
\begin{eqnarray}
\Lambda \left( r,\varsigma ,m,n\right) &=&\frac{1}{\sqrt{m!}}\left( -i\sqrt{%
\frac{\sinh r\cosh r}{2}}\right) ^{n}\left. \frac{d^{m}}{d\alpha ^{m}}%
\left\langle 0\right\vert H_{n}\left( i\sqrt{\frac{\cosh r}{2\sinh r}}%
a\right) \int \frac{d^{2}z}{\pi }\left\vert z\right\rangle \left\langle
z\right\vert S\left( \nu \right) \left\vert \alpha \right\rangle \right\vert
_{\alpha =0}  \notag \\
&=&\frac{1}{\sqrt{m!\cosh \nu }}\left( -i\sqrt{\frac{\sinh r\cosh r}{2}}%
\right) ^{n}\left. \frac{d^{m}}{d\alpha ^{m}}\exp \left( -\frac{\tanh \nu }{2%
}\alpha ^{2}\right) \right.  \label{ac2} \\
&&\times \left. \int \frac{d^{2}z}{\pi }H_{n}\left( i\sqrt{\frac{\cosh r}{%
2\sinh r}}z\right) \exp \left[ -\left\vert z\right\vert ^{2}+\frac{\tanh \nu 
}{2}z^{\ast 2}+z^{\ast }\alpha \sec h\nu \right] \right\vert _{\alpha =0}, 
\notag
\end{eqnarray}%
from (\ref{aa10}), then it follows%
\begin{eqnarray}
\Lambda \left( r,\varsigma ,m,n\right) &=&\frac{1}{\sqrt{m!\cosh \nu }}%
\left( -i\sqrt{\frac{\sinh r\cosh r}{2}}\right) ^{n}\left. \frac{d^{m}}{%
d\alpha ^{m}}\exp \left( -\frac{\tanh \nu }{2}\alpha ^{2}\right) \right. 
\notag \\
&&\times \left. \frac{\partial {}^{n}}{\partial t{}^{n}}\exp \left(
-t^{2}\right) \int \frac{d^{2}z}{\pi }\left. \exp \left( -\left\vert
z\right\vert ^{2}+\frac{\tanh \nu }{2}z^{\ast 2}+z^{\ast }\alpha \sec h\nu
+2i\sqrt{\frac{\cosh r}{2\sinh r}}zt\right) \right\vert _{t=0}\right\vert
_{\alpha =0}.  \label{ac3}
\end{eqnarray}%
Thus Eq.(\ref{aa12}) can tell us%
\begin{eqnarray}
\Lambda \left( r,\varsigma ,m,n\right) &=&\frac{1}{\sqrt{m!\cosh \nu }}%
\left( -i\sqrt{\frac{\sinh r\cosh r}{2}}\right) ^{n}\left. \frac{d^{m}}{%
d\alpha ^{m}}\exp \left( -\frac{\tanh \nu }{2}\alpha ^{2}\right) \right. 
\notag \\
&&\times \left. \left. \frac{\partial {}^{n}}{\partial t{}^{n}}\exp \left[
-\left( 1+\frac{\tanh \nu }{\tanh r}\right) t^{2}+it\alpha \sqrt{\frac{%
2\cosh r}{\sinh r}}\sec h\nu \right] \right\vert _{t=0}\right\vert _{\alpha
=0}.  \label{ac4}
\end{eqnarray}%
Using (\ref{aa10}) again, we can see%
\begin{eqnarray}
\Lambda \left( r,\varsigma ,m,n\right) &=&\frac{1}{\sqrt{m!\cosh \nu }}%
\left( -i\sqrt{\frac{\sinh r\cosh r}{2}}\right) ^{n}\left( 1+\frac{\tanh \nu 
}{\tanh r}\right) ^{\frac{n}{2}}  \notag \\
&&\times \left. \frac{d^{m}}{d\alpha ^{m}}H_{n}\left( \frac{i\alpha \sec
h\nu }{\sqrt{2\left( \tanh r+\tanh \nu \right) }}\right) \exp \left( -\frac{%
\tanh \nu }{2}\alpha ^{2}\right) \right\vert _{\alpha =0},  \label{ac5}
\end{eqnarray}%
then it follows%
\begin{eqnarray}
\Lambda \left( r,\varsigma ,m,n\right) &=&\frac{1}{\sqrt{m!\cosh \nu }}%
\left( -i\sqrt{\frac{\sinh r\cosh r}{2}}\right) ^{n}\left( 1+\frac{\tanh \nu 
}{\tanh r}\right) ^{\frac{n}{2}}  \notag \\
&&\times \left. \sum_{k=0}^{m}\dbinom{m}{k}\frac{d^{k}}{d\alpha ^{k}}H_{n}%
\left[ \frac{i\alpha \sec h\nu }{\sqrt{2\left( \tanh r+\tanh \nu \right) }}%
\right] \frac{d^{m-k}}{d\alpha ^{m-k}}\exp \left( -\frac{\tanh \nu }{2}%
\alpha ^{2}\right) \right\vert _{\alpha =0}.  \label{ac6}
\end{eqnarray}%
From (\ref{aa16}), we have%
\begin{eqnarray}
\Lambda \left( r,\varsigma ,m,n\right) &=&\frac{1}{\sqrt{m!\cosh \nu }}%
\left( -i\sqrt{\frac{\sinh r\cosh r}{2}}\right) ^{n}\left( 1+\frac{\tanh \nu 
}{\tanh r}\right) ^{\frac{n}{2}}\sum_{k=0}^{m}\dbinom{m}{k}\frac{2^{k}n!}{%
\left( n-k\right) !}  \notag \\
&&\times \left( \frac{i\sec h\nu }{\sqrt{2\left( \tanh r+\tanh \nu \right) }}%
\right) ^{k}\left. H_{n-k}\left( \frac{i\alpha \sec h\nu }{\sqrt{2\left(
\tanh r+\tanh \nu \right) }}\right) \frac{d^{m-k}}{d\alpha ^{m-k}}\exp
\left( -\frac{\tanh \nu }{2}\alpha ^{2}\right) \right\vert _{\alpha =0}
\label{ac7}
\end{eqnarray}%
and due to $\left. \frac{d^{m-k}}{d\alpha ^{m-k}}\exp \left( -\frac{\tanh
\nu }{2}\alpha ^{2}\right) \right\vert _{\alpha =0}=\left( \frac{\tanh \nu }{%
2}\right) ^{\frac{m-k}{2}}H_{m-k}\left( 0\right) $, finally we can derive%
\begin{eqnarray}
\Lambda \left( r,\varsigma ,m,n\right) &=&\frac{1}{\sqrt{m!\cosh \left(
\varsigma -r\right) }}\left( -i\sqrt{\frac{\sinh 2r}{4}}\right) ^{n}\left[ 1+%
\frac{\tanh \left( \varsigma -r\right) }{\tanh r}\right] ^{\frac{n}{2}}\left[
\frac{\tanh \left( \varsigma -r\right) }{2}\right] ^{\frac{m}{2}}  \notag \\
&&\times \sum_{k=0}^{m}\dbinom{m}{k}\frac{\left( 2i\right) ^{k}n!}{\left(
n-k\right) !}\left[ \frac{2}{\left( \tanh r+\tanh \left[ \varsigma -r\right]
\right) \sinh \left( 2\varsigma -2r\right) }\right] ^{\frac{k}{2}%
}H_{n-k}\left( 0\right) H_{m-k}\left( 0\right) .  \label{ac8}
\end{eqnarray}

\end{document}